\documentclass[journal]{IEEEtran}
\usepackage[T1]{fontenc}
\usepackage[latin9]{inputenc}
\usepackage{color}
\usepackage{array}
\usepackage{bm}
\usepackage{multirow}
\usepackage{amsmath}
\usepackage{amsthm}
\usepackage{amssymb}
\usepackage{graphicx}
\usepackage[bookmarks=true,bookmarksnumbered=true,bookmarksopen=true,bookmarksopenlevel=1,
 breaklinks=false,pdfborder={0 0 1},backref=false,colorlinks=false]
 {hyperref}
\hypersetup{pdftitle={Your Title},
 pdfauthor={Your Name},
 pdfborderstyle=,pdfpagelayout=OneColumn,pdfnewwindow=true,pdfstartview=XYZ,plainpages=false}

\makeatletter

\providecommand{\tabularnewline}{\\}

\theoremstyle{plain}
\newtheorem{thm}{\protect\theoremname}
\theoremstyle{plain}
\newtheorem{lem}[thm]{\protect\lemmaname}

\providecommand{\lemmaname}{Lemma}
\providecommand{\theoremname}{Theorem}
\usepackage{cite}
\usepackage{hyperref}
\hypersetup{
  colorlinks=true,   
  linkcolor=blue,    
  citecolor=blue,   
  urlcolor=blue      
}
\usepackage{diagbox}
\usepackage{multirow}
\usepackage{array}

\ifdefined\showcaptionsetup
 \PassOptionsToPackage{caption=false}{subfig}
\fi
\usepackage{subfig}
\makeatother

\providecommand{\lemmaname}{Lemma}
\providecommand{\theoremname}{Theorem}

\begin{document}
\title{Information-Preserving CSI Feedback: Invertible Networks with Endogenous
Quantization and Channel Error Mitigation}
\author{Haotian~Tian,~\IEEEmembership{Student~Member,~IEEE},~Lixiang~Lian,~\IEEEmembership{Member,~IEEE},
Jiaqi Cao, Sijie Ji,~\IEEEmembership{Member,~IEEE}\thanks{\emph{(Corresponding author: Lixiang Lian.)}}\thanks{Haotian~Tian, Lixiang Lian and Jiaqi Cao are with the School of Information
Science and Technology, ShanghaiTech University, Shanghai 201210,
China (e-mail: \{tianht2022, lianlx, caojq\}@shanghaitech.edu.cn).}\thanks{Sijie Ji is a Schmidt Science Fellow with Division of Engineering
and Applied Science, California Institute of Technology, CA 91125,
USA (email: sijieji@caltech.edu).}}
\maketitle
\begin{abstract}
Deep learning has emerged as a promising solution for efficient channel
state information (CSI) feedback in frequency division duplex (FDD)
massive MIMO systems. Conventional deep learning-based methods typically
rely on a deep autoencoder to compress the CSI, which leads to irreversible
information loss and degrades reconstruction accuracy. This paper
introduces InvCSINet, an information-preserving CSI feedback framework
ba\textcolor{black}{sed on invertible neural networks (INNs). }By
leveraging the bijective nature of INNs, the model ensures information-preserving
compression and reconstruction with shared model parameters. To address
practical challenges such as quantization and channel-induced errors,
we endogenously integrate an adaptive quantiza\textcolor{black}{tion
mo}dule, a differentiable bit-channel distortion module and an information
compensation module into the INN architecture. This design enables
the network to learn and compensate the information loss during CSI
compression, quantization, and noisy transmission, thereby preserving
the CSI integrity throughout the feedback process. Simulation results
validate the effectiveness of the proposed scheme, demonstrating superior
CSI recovery performance and robustness to practical impairments with
a lightweight architecture.
\end{abstract}


\IEEEpeerreviewmaketitle{}

\section{Introduction}

\IEEEPARstart{M}{assive} Multiple Input Multiple Output (MIMO) is
a cornerstone of next-generation mobile communications, utilizing
large-scale antenna arrays and beamforming to enhance wireless communication
performances \cite{marzetta2010noncooperative,hoydis2013massive}.
In frequency division duplexing (FDD) systems, the non-reciprocal
nature of channel state information (CSI) necessitates feedback from
the user equipment (UE) to the base station (BS) for downlink transmission
designs. As the number of antennas increases, CSI feedback overhead
becomes prohibitively large, making efficient CSI feedback a key challenge
in FDD massive MIMO systems.

Traditional CSI feedback schemes rely on codebook \cite{alevizos2018limited,love2008overview,kang2021novel}
or compressed sensing (CS) \cite{liu2018downlink,wei2022joint,li2024downlink}
techniques to reduce the feedback overhead. However, codebook-based
approaches suffer from low resolution for accurate CSI feedback. CS-based
methods exploit channel sparsity to recover CSI from few-bit measurements.
However, their performance degrades significantly when the sparsity
assumption is violated or when the measurements are severely compressed.
Deep learning (DL) has emerged as an effective tool for CSI feedback
problem, providing higher accuracy, stronger adaptability, and scalability
\cite{guo2022overview}. Conventional DL-based CSI feedback schemes
typically adopt deep autoencoder (DAE) architectures for CSI compression
and reconstruction \cite{wen2018deep,ji2021CLnet,chen2019novel,xu2022deep,lu2020multi,hu2023csi,guo2024bayesian,ye2020deep,chang2021deep,ravula2021deep,wang2018deep,lu2022binarized,guo2020convolutional,liang2022changeable,zhang2023quantization,liu2020efficient,cao2022information,cao2023adaptive}.
However, such hard compression inherently leads to information loss,
rendering the reconstruction process ill-posed and compromising accuracy.
Moreover, existing approaches often o\textcolor{black}{verlook the
combined distortions introduced by quantization and wireless transmission
errors during feedback, particularly in the model learning phase.
The quantization operation is non-differentiable, making it difficult
to incorporate into gradient-based learning. Similarly, channel-induced
errors, especially in the transmission of digital signals, are hard
to model in a form compatible with neural network optimization. Consequently,
most existing approaches avoid jointly considering quantization and
noisy bit-channels during training, leading to a mismatch between
model design and real-world transmission conditions.}

\subsection{Related Works}

\subsubsection{DAE-based CSI Feedback}

DAE-based CSI feedback schemes have been proposed and shown to significantly
outperform traditional approaches. For instance, the networks in \cite{wen2018deep,chen2019novel,lu2020multi,hu2023csi,guo2024bayesian,ye2020deep,chang2021deep,xu2022deep}
were designed based on DAE directly, such as CLNet\cite{ji2021CLnet},
CsiNet \cite{wen2018deep}, CsiNet+\cite{chen2019novel,xu2022deep},
CRNet\cite{lu2020multi}, CSI-StriperFormer\cite{hu2023csi}, Hierarchical
Sparse AE \cite{guo2024bayesian}, DNNet \cite{ye2020deep}, ATNet
\cite{chang2021deep}. Some works incorporated other types of neural
networks into the DAE to leverage additional information. For example,
\cite{wang2018deep} utilized Long Short-Term Memory (LSTM) networks
to exploit the temporal correlations of CSI. Although DAE has been
widely adopted for CSI feedback, its effectiveness in practical systems
is contrained. First, the DAE encoder performs a hard compression
via a general neural network, which is typically non-invertible. This
leads to the loss of cricial information in the CSI, thereby compromising
the reconstruction accuracy. Second, the encoder and decoder are implemented
using separate large networks, as the DAE treats compression and reconstruction
as two entirely distinct processes. The large model size of DAE results
in increased storage cost and reduced training efficiency.

\subsubsection{Quantization-Aware CSI Feedback}

In CSI feedback, the uplink channel only supports bit-level transmission,
making quantization of compressed features necessary. However, the
non-differentiability of quantization poses challenges for gradient-based
DL training. Existing methods can be categorized into multi-step training
\cite{chen2019novel,guo2020convolutional,zhang2023quantization},
gradient approximation \cite{guo2024bayesian,ravula2021deep,lu2022binarized,liang2022changeable},
and soft-to-hard quantizers \cite{liu2020efficient}. The multi-step
approach first trains encoder and decoder without quantization, then
optimizes a standalone dequantizer, followed by joint fine-tuning,
which compromises accuracy due to suboptimal separation of stages.
Gradient approximation methods use identity mappings \cite{lu2022binarized}
or hyperparameterized surrogates \cite{liang2022changeable}, but
often suffer from gradient mismatch. Some works \cite{guo2024bayesian,ravula2021deep}
added uniformly distributed noise to latent features to enable differentiability,
but fail to preserve the discrete nature of quantization, causing
data distribution mismatches between training and testing. Soft-to-hard
quantizers, such as \cite{liu2020efficient}, approximate quantization
using differentiable functions (e.g., sigmoid combinations), allowing
end-to-end training with controllable approximation accuracy and better
consistency. Nevertheless, existing schemes generally lack adaptive
quantization (AQ) mechanisms to optimize quantization points based
on data distribution, and often ignore channel noise during digital
transmission, as quantization is typically omitted during training.
Both factors limit CSI recovery performance under practical conditions.

\subsubsection{Channel-Error-Aware CSI Feedback}

Several works have addressed CSI feedback through imperfect uplink
channels \cite{ye2020deep,chang2021deep,xu2022deep}. For example,
\cite{ye2020deep} introduced a noise extraction unit (NEU) to denoise
the received codewords at the BS. However, the quantization of latent
features were not considered. \cite{chang2021deep} deployed a bit
error correction block (ECBlock) before the decoder at the BS to compensate
the uplink channel errors , which required a multi-step training.
Moreover, a straight-through estimator was used in \cite{ye2020deep}
for quantization, which can lead to significant gradient mismatch.
\cite{xu2022deep} applied a Deep Joint Source-Channel Coding (JSCC)
system to the CSI feedback task, incorporating channel error in the
forward pass, while the effects of quantization and channel noise
were not accounted for during network training.

Many existing studies address either quantization-aware or channel-error-aware
CSI feedback, overlooking their joint impact during the algorithm
design. This stems from two key challenges. First, after quantization,
the uplink channel supports only bit transmission, where channel noise
can flip bits with a certain probability, making the whole process
inherently non-differentiable. Second, both channel noise and quantization
errors are added to the compressed features, posing challenges for
eliminating mixed distortions and accurately reconstructing the CSI.

\subsection{Our Contributions}

This paper presents InvCSINet, an information-preserving CSI feedback
framework built upon invertible neural networks (INNs). By embedding
a differentiable adaptive quantization (DAQ) module, a differentiable
bit-channel distortion (D-BCD) module and an information compensation
(IC) module within the INN architecture, the proposed InvCSINet enables
the network to learn and compensate for information loss arising from
compression, quantization, and transmission noise, thereby maintaining
CSI integrity throughout the feedback process. The main contributions
are summarized as follows:
\begin{itemize}
\item \textbf{INN-Based CSI Feedback: }Compared to traditional neural networks,
INNs ensure strict bijectivity between input and output, with the
encoder corresponding to the forward pass and the decoder to the inverse.
By regulating the forward and backward processes of INN, we demonstrate
that the INN-based CSI feedback can faithfully reconstruct the true
distribution of the original CSI at the decoder output.
\item \textbf{INN-Based Robust CSI Feedback with Quantization and Channel
Errors}: In practical scenarios, the decoder receives a quantization-channel
joint distorted version of the encoder's output. To preserve information
throughout the entire CSI feedback process under these impairments,
the proposed InvCSINet contains three key designs. 1)\textbf{ DAQ
Module:} By learning the latent feature distribution, the DAQ module
adaptively designs non-uniform quantization schemes, thereby minimizing
information loss under limited bit budgets and supporting end-to-end
optimization. 2)\textbf{ D-BCD Module:} By modeling the effect of
channel errors using Gumbel-Softmax techniques, D-BCD module enables
the network to naturally account for bit-level errors during training,
thereby mitigating their adverse impact on CSI feedback accuracy.
3) \textbf{IC Module: }IC module is introduced to mitigate the distortion
caused by quantization and channel errors, which consists of a Latent
Alignment Network (LAN) and an auxiliary variable learning module.
We demonstrate that, when trained to convergence with zero loss, InvCSINet
preserves the true distribution of the original CSI, ensuring distributional
fidelity at the output.
\item \textbf{Experimental Verification:} Simulation results show that the
proposed InvCSINet achieves better CSI feedback accuracy under limited
feedback overhead compared to other network architectures. By inherently
embedding quantization and channel-induced distortion into the network,
the proposed scheme demonstrates substantial performance gains in
practical scenarios.
\end{itemize}
The rest of paper is organized as follows. In Section \ref{sec_2},
we present the signal model and problem formulation of CSI feedback
in FDD systems. Section \ref{sec:Invertible-CSI-Feedback} presents
the INN-based CSI feedback framework with ideal transmission conditions.
Section \ref{sec:InvCSINet:-Robust-CSI} extends this framework by
incorporating the quantization effect and bit-level channel errors,
leading to the proposed InvCSINet. Finally, the extensive simulation
results are provided in Section \ref{sec_4} and conclusions are drawn
in Section \ref{sec_5}\textit{.}

\section{System Model and Problem Formulation\label{sec_2}}

\subsection{CSI Feedback System Model}

Consider an Orthogonal Frequency Division Multiplexing (OFDM) system
with $N_{c}$ subcarriers, where the BS is equipped with $N_{t}\gg1$
uniform linear array (ULA) antennas, and the user is equipped with
$N_{r}$ ULA antennas. At the $n$-th subcarrier, the downlink CSI
matrix that needs to be fed back from user to BS is denoted as $\mathbf{H}_{n}\in\mathbb{C}^{N_{r}\times N_{t}}$.
The collection of CSI matrices within the whole bandwidth is $\mathbf{H}=\begin{bmatrix}\mathbf{H}_{1},\dots,\mathbf{H}_{N_{c}}\end{bmatrix}\in\mathbb{C}^{N_{r}\times N_{t}N_{c}}$.
Due to the large number of antennas at both the BS and user side,
and the wide bandwidth in OFDM systems, the overall dimension of the
complex CSI $\mathbf{H}$ (i.e., $N=2N_{r}N_{t}N_{c})$ to be fed
back can be prohibitively high. Therefore, efficient feedback strategies
are essential to reduce overhead while maintaining reconstruction
accuracy.

\subsection{Problem Formulation}

In practical digital transmission systems, the CSI feedback process
can be formulated as follows. The user compresses the original CSI
matrix $\mathbf{H}$ into a low-dimensional latent feature $\mathbf{z}$
through an encoder $f$ parameterized by $\theta_{f}$, i.e., $\mathbf{z}=f(\mathbf{H};\theta_{f})$,
where the dimension of $\mathbf{z}$ is denoted as $M$. The latent
feature $\mathbf{z}$ is then quantized by a quantizer $Q(\cdot)$,
after which, the feature is discretized in both magnitude and time
scale. The impact of noisy bit-level transmission is modeled by a
channel distortion function $\varOmega(\cdot)$, which jointly characterizes
the entire process that the quantized output undergoes before recovered
at the BS. This includes bitstream encoding, modulation, transmission
through a noisy channel, as well as demodulation, detection, and decoding
at the BS. The recovered representation at the BS can be expressed
as $\hat{\mathbf{z}}=\Omega\left(Q(\mathbf{z})\right)$. At the BS,
a decoder $g$ parameterized by $\theta_{g}$ is used to reconstruct
the original CSI from the corrupted feature, i.e., $\widehat{\mathbf{H}}=g(\hat{\mathbf{z}};\theta_{g})$.
The objective of CSI feedback is to design the encoder, quantizer,
and decoder under limited feedback constraint of $MB$ bits, such
that the reconstruction error is minimized, i.e., 
\begin{align}
\min & ~\ \mathbb{E}_{\mathbf{H}}\left[\left\Vert \widehat{\mathbf{H}}-\mathbf{H}\right\Vert _{F}^{2}\right]\label{eq: problem_formualtion}\\
\text{s.t.} & ~\ \mathrm{Feedback\ overhead}\leq MB\text{\ bits,}\nonumber 
\end{align}
where $B$ stands for the transmission bit budget per feature dimension.

However, because the quantizer and bit-channel distortion are difficult
to model as differentiable processes, it remains challenging for the
network to endogenously account for their effects during the optimization
of transmission schemes. Most existing CSI feedback schemes neglect
the simultaneous distortions induced by quantization and channel errors.
In Section \ref{sec:Invertible-CSI-Feedback}, we first consider an
ideal feedback scenario, where quantization and channel errors are
ignored, i.e., $\Omega\left(Q(\cdot)\right)=\mathbf{I}$, to introduce
the proposed INN-based CSI feedback scheme. In Section \ref{sec:InvCSINet:-Robust-CSI},
we extend our design to practical scenarios by endogenously integrating
a DAQ module, a D-BCD module and an IC module into the INN architecture.
This integration allows the network to adaptively learn to compensate
for compression and transmission impairments during training, thereby
improving the robustness and accuracy of CSI feedback.

\section{Invertible CSI Feedback Scheme with Ideal Transmission\label{sec:Invertible-CSI-Feedback}}

\begin{figure}
\subfloat[Super Resolution Network.\label{fig:superresolution}]{\begin{centering}
\includegraphics[viewport=240bp 445bp 780bp 545bp,clip,scale=0.47]{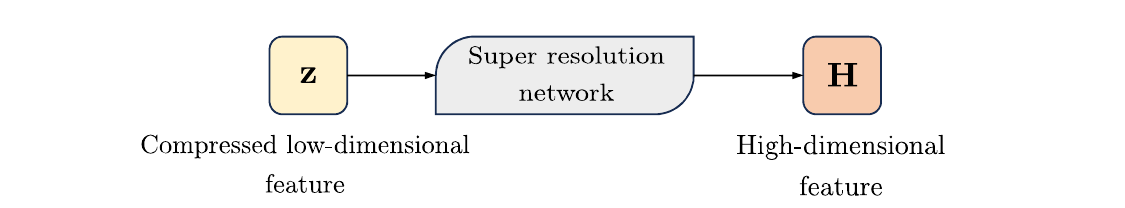}
\par\end{centering}
}

\subfloat[End-to-end Encoder.\label{fig:DAE-based}]{\begin{centering}
\includegraphics[viewport=240bp 300bp 780bp 400bp,clip,scale=0.47]{img/Conventions}
\par\end{centering}
}\caption{Conventional solutions for CSI feedback. \label{fig:Conventional-solutions-for}}
\end{figure}

In this section, we illustrate how to implement an invertible CSI
feedback process under ideal transmission conditions, where quantization
and channel-induced errors are omitted. Since the dimension of $\mathbf{z}$
is significantly smaller than that of the original CSI $\mathbf{H}$,
i.e., $M\ll N$, the compression process inevitably causes information
loss, rendering the recovery task from $\mathbf{z}$ to $\mathbf{H}$
ill-posed. To address this issue, conventional solutions leverage
supervised DL techniques for CSI recovery. For example, as shown in
Fig. \ref{fig:superresolution}, a super-resolution neural network
deployed at BS learns a direct mapping from the low-resolution features
$\mathbf{z}$ to high-resolution CSI $\mathbf{H}$ using a large amount
of training data, without modeling the encoding process. Alternatively,
Fig. \ref{fig:DAE-based} adopts a DAE framework, in which the encoder
compresses the CSI at the user side, and the decoder at the BS attempts
to reconstruct the CSI from the received low-dimensional features.
While this end-to-end learning approach offers better consistency
between compression and reconstruction but still suffers from information
loss due to limited feature capacity. To address this limitation,
we propose an INN-based CSI feedback framework that enables information
preservation under ideal transmission conditions and provides a principled
mechanism to compensate for information loss at the decoder when quantization
and channel errors are introduced. In the following, we first introduce
the overall framework, followed by a detailed description of the INN
architecture, the bi-directional training strategy, and a discussion
of the fundamental principles underlying the INN design.

\subsection{Overall Framework}

\begin{figure}[t]
\begin{centering}
\includegraphics[viewport=90bp 85bp 780bp 510bp,clip,scale=0.37]{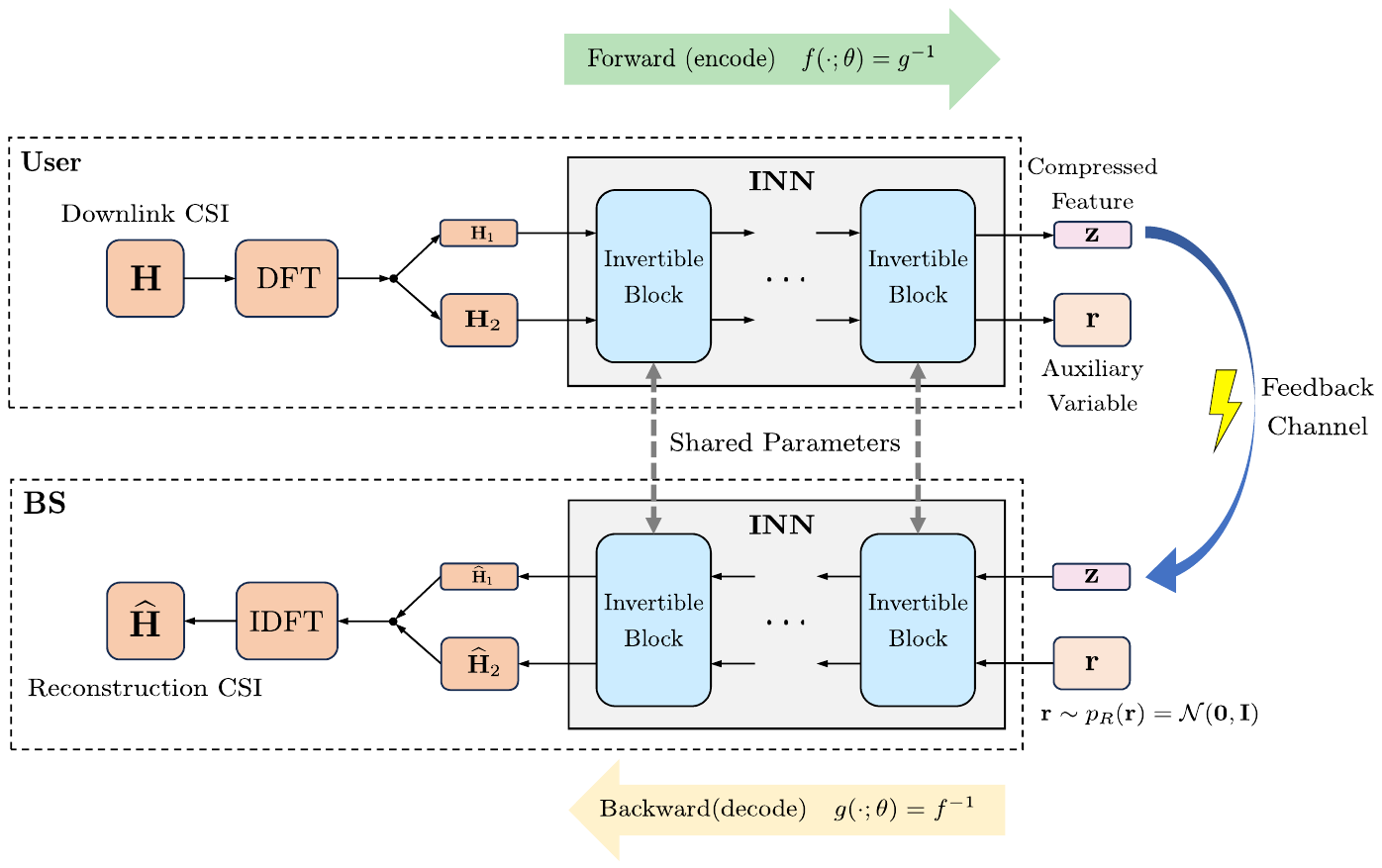}
\par\end{centering}
\caption{The overall framework of INN-based CSI feedback with ideal transmission.\label{Inv_scheme}}
\end{figure}

The information-preserving principle of the proposed INN-based CSI
feedback is illustrated in Fig. \ref{Inv_scheme}. Specifically, the
high-dimensional CSI $\mathbf{H}$ is encoded by a bijective mapping
$f(\cdot;\theta)$, yielding a compressed feature $\mathbf{z}$ and
an auxiliary feature $\mathbf{r}$, where $\mathbf{r}$ captures the
information in $\mathbf{H}$ that is not contained in $\mathbf{z}$,
i.e. the lossy information during feedback. Due to the invertibility
of $f$, the original CSI $\mathbf{H}$ can be perfectly reconstructed
by applying the inverse mapping $g(\cdot;\theta)=f^{-1}(\cdot;\theta)$
to the pair $(\mathbf{z},\mathbf{r})$. However, in the CSI feedback
scenarios, only the compressed feature $\mathbf{z}$ is transmitted
to the BS, while the auxiliary feature $\mathbf{r}$ is discarded
and unavailable at the decoder. Therefore, it is crucial to design
an auxiliary feature that enables effective compensation at decoder
despite the absence of $\mathbf{r}$.

To address this, at the BS, the auxiliary variable $\mathbf{r}$ is
draw from a multi-variate standard normal distribution, i.e., $\mathbf{r}\sim p_{R}(\mathbf{r})=\mathcal{N}(\mathbf{r};\bm{0},\mathbf{I})$,
following the practice in \cite{ardizzone2018analyzing}. After sampling
according to $p_{R}$, a sample $\mathbf{r}$ is combined with the
received $\mathbf{z}$ for CSI reconstruction via the decoder $g(\cdot;\theta)$.
Both the encoder $f$ and decoder $g$ share the same set of parameters
$\theta$ and are jointly learned to satisfy the following relationships:
\begin{align}
\mathrm{Forward:}\left[\mathbf{z,r}\right] & =f(\mathbf{H};\theta)=\left[f_{\mathbf{z}}(\mathbf{H};\theta),f_{\mathbf{r}}(\mathbf{H};\theta)\right]\\
 & =g^{-1}(\mathbf{H};\theta),\nonumber 
\end{align}
\begin{equation}
\mathrm{Backward:}\mathbf{H}=g(\mathbf{z},\mathbf{r};\theta),\mathrm{with}\ \mathbf{r}\sim p_{R}(\mathbf{\mathbf{r}})=\mathcal{N}(\mathbf{r};\bm{0},\mathbf{I}).
\end{equation}
The bijective relation of $f=g^{-1}$ is enforced by the INN architecture,
provided that the dimension of input matches that of output. Because
the dimension of $\mathbf{H}$ is $N$, and the dimension of latent
feature $\mathbf{z}$ is $M$, to guarantee the bijectivity, the dimension
of auxiliary variable $\mathbf{r}$ must be $N-M$.

The information-preserving property of the INN architecture offers
inherent performance guarantees for CSI recovery. Furthermore, its
support for bidirectional training enables more efficient model optimization.
Sharing parameters between the encoder and decoder further reduces
model complexity, enhancing training efficiency, and lowers deployment
overhead.

\subsection{Invertible Architecture}

\begin{figure}
\begin{centering}
\includegraphics[viewport=190bp 250bp 570bp 400bp,clip,scale=0.5]{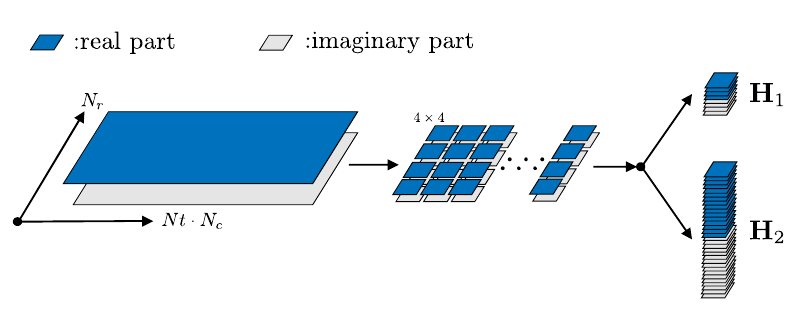}
\par\end{centering}
\caption{The segmentation strategy.\label{fig:The-segmentation-strategy}}
\end{figure}
\begin{figure}[t]
\subfloat[Forward process.\label{block_a-1}]{\begin{centering}
\includegraphics[viewport=90bp 320bp 770bp 520bp,clip,scale=0.36]{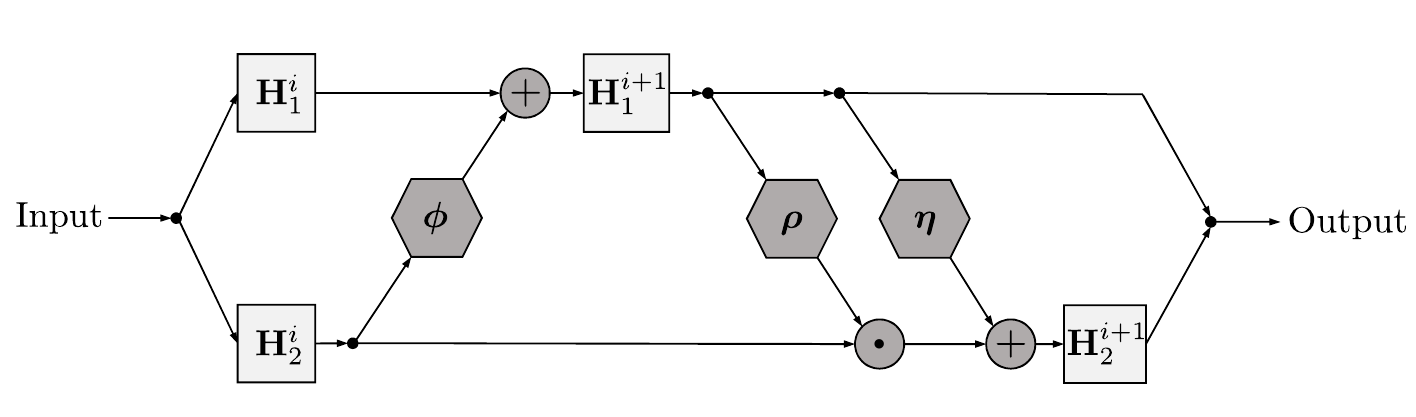}
\par\end{centering}
}

\subfloat[Backward process.\label{block_b-1}]{\begin{centering}
\includegraphics[viewport=80bp 105bp 760bp 304bp,clip,scale=0.36]{img/Invertible_Block}
\par\end{centering}
}\caption{The architecture of an invertible block.\label{Invertible Block}}
\end{figure}

Conventional neural networks do not inherently guarantee bijectivity.
Constructing a bijective function requires specific architectural
constraints. In this paper, we construct the fully invertible network
following the architecture in \cite{dinh2016density}, which is composed
of a sequence of invertible blocks. As illustrated in Fig. \ref{Inv_scheme},
before being fed into the INN architecture, CSI $\mathbf{H}$ is first
transformed into the angular domain via 2D Discrete Fourier Transform
(DFT). This transformation reveals more explicit structural patterns
in the CSI, facilitating more effective compression. Moreover, the
DFT is invertible, ensuring information lossless during transformation.
After applying the DFT, the angular-domain CSI is converted into its
real-valued representation by concatenating the real and imaginary
components. Subsequently, the real-valued CSI matrix is partitioned
into two segments, $\mathbf{H}_{1}$ and $\mathbf{H}_{2}$, based
on the target compression ratio. The dimension of $\mathbf{H}_{1}$
matches that of the compressed latent variable $\mathbf{z}$, while
the dimension of $\mathbf{H}_{2}$ is consistent with that of the
auxiliary variable $\mathbf{r}$. The segmentation strategy is illustrated
in Fig. \ref{fig:The-segmentation-strategy}, where we first reshape
the complex channel data $\mathbf{H}$ by stacking its real part and
imaginary part along the channel dimension, then we segment it along
the height and the width dimension to generate patches of size $4\times4$.
Finally, $\mathbf{H}_{1}$ and $\mathbf{H}_{2}$ are obtained by splitting
along the channel dimension. After segmentation, $\mathbf{H}_{1}$
and $\mathbf{H}_{2}$ are jointly processed by the sequence of invertible
blocks.

Each invertible block consists of two affine coupling layers. The
structure of the $i$-th block is illustrated in Fig. \ref{Invertible Block}.
In particular, for the $i$-th invertible block, it receives $\mathbf{H}_{1}^{i}$
and $\mathbf{H}_{2}^{i}$ as input, and outputs $\mathbf{H}_{1}^{i+1}$
and $\mathbf{H}_{2}^{i+1}$ through the following operation:
\begin{align}
\mathbf{H}_{1}^{i+1}= & \mathbf{H}_{1}^{i}+\bm{\phi}(\mathbf{H}_{2}^{i})\\
\mathbf{H}_{2}^{i+1}= & \mathbf{H}_{2}^{i}\odot\exp\big[\bm{\rho}(\mathbf{H}_{1}^{i+1})\big]+\bm{\eta}(\mathbf{H}_{1}^{i+1}).
\end{align}
Given the output $\mathbf{H}_{1}^{i+1}$ and $\mathbf{H}_{2}^{i+1}$,
the inverse process can be easily derived as
\begin{align}
\mathbf{H}_{1}^{i}= & \mathbf{H}_{1}^{i+1}-\bm{\phi}(\mathbf{H}_{2}^{i})\\
\mathbf{H}_{2}^{i}= & \big[\mathbf{H}_{2}^{i+1}-\bm{\eta}(\mathbf{H}_{1}^{i+1})\big]\oslash\exp\big[\bm{\rho}(\mathbf{H}_{1}^{i+1})\big],
\end{align}
where $\odot$, and $\oslash$ are element-wise multiplication, and
division respectively. There are natural constant exponent operations
in multiplication and division to avoid division by zero. $\bm{\phi}$,
$\bm{\rho}$, $\bm{\eta}$ do not require any invertibility and can
be any complex network, such as fully connected neural networks (FCNNs),
convolutional neural networks (CNNs), etc. Permulation layers can
be inserted between two invertible blocks to shuffle the input elements
of the following layer. This can introduce different combinations
of $[\mathbf{H}_{1}^{i},\mathbf{H}_{2}^{i}]$ across blocks to enhance
interactions among the individual variables \cite{kingma2018glow}.

\subsection{Bi-Directional Training of INN}

The INN architecture enables simultaneous regularization of the network
parameters for both the forward and backward processes, leading to
more effective training. Specifically, we alternately perform forward
and backward passes, accumulate gradients from both directions based
on their respective losses, and subsequently update the model parameters.
The loss function comprises two components, as illustrated in the
following.

\subsubsection{Forward Loss}

For the forward pass, we regularize the distribution of the latent
and auxiliary variables by minimizing the divergence between the joint
distribution of the network outputs $q(\mathbf{z}=f_{\mathbf{z}}(\mathbf{H};\theta),\mathbf{r}=f_{\mathbf{r}}(\mathbf{H},\theta))$
with the product of the marginal distribution $p(\mathbf{z}=f_{\mathbf{z}}(\mathbf{H};\theta))$
and the prior distribution $p_{R}(\mathbf{r})$ through a loss $\mathcal{L}_{\mathbf{r}}$,
given by 
\begin{equation}
\mathcal{L}_{\mathbf{r}}(q(\mathbf{z},\mathbf{r}),p(\mathbf{z})p_{R}(\mathbf{r}))=\mathrm{MMD}^{2}(q(\mathbf{z},\mathbf{r}),p(\mathbf{z})p_{R}(\mathbf{r})),\label{eq:loss_DistributionMatch}
\end{equation}
where $\text{\ensuremath{\mathcal{L}_{\mathbf{r}}}}$ is implemented
through Maximum Mean Discrepancy (MMD) \cite{gretton2012kernel}.
MMD can be used to measure the discrepancy between two distributions
based on data samples, whose calculation will be introduced later.

During back-propagation, we block the gradients of $\mathcal{L}_{r}$
with respect to $\mathbf{z}$ to make sure that the update of the
network only affects the predictions of $\mathbf{r}$ and do not affect
the predictions of $\mathbf{z}$. The loss function $\mathcal{L}_{\mathbf{r}}$
enforces that the distribution of the auxiliary variable $\mathbf{r}$,
produced by forward process, conforms to the predefined normal distribution
$p_{R}$. Moreover, it promotes statistical independence between the
forward outputs $\mathbf{r}$ and $\mathbf{z}$. This independence
is essential. Firstly, it minimizes the information redundancy between
$\mathbf{r}$ and $\mathbf{z}$, thereby improving compression efficiency.
Secondly, it ensures that the backward process can accurately reconstruct
the original input by matching the prior distribution of $\mathbf{H}$,
which can be theoretically guaranteed through the following theorem.
Before presenting the main theorem, we first introduce the following
lemma, which highlights the favorable distribution-preserving property
of the INN.
\begin{lem}
[Ditribution Preservation of INN \cite{ardizzone2018analyzing}]\label{lem:Ditribution-Preservation-of}If
some bijective function $f:x\rightarrow z$ transforms a probability
density $p_{X}(x)$ to $p_{Z}(z)$, then the inverse function $f^{-1}$
transforms $p_{Z}(z)$ back to $p_{X}(x)$.
\end{lem}
\begin{thm}
[Ditributional Matching of INN-Based CSI Feedback] \label{thm:pdf_match}Let
$p_{H}(\mathbf{H})$ denote the prior distribution of the CSI. Suppose
an INN $f(\mathbf{H})=[\mathbf{z},\mathbf{r}]$ is trained to minimize
the unsupervised loss $\mathcal{L}_{\mathbf{r}}=\mathrm{MMD}^{2}(q(\mathbf{z},\mathbf{r}),p(\mathbf{z})p_{R}(\mathbf{r}))$,
where $p_{R}(\mathbf{r})=\mathcal{N}(\mathbf{r};\bm{0},\mathbf{I})$,
and assume the loss reaches zero. Then, for any input $\mathbf{z}=f_{\mathbf{z}}(\mathbf{H};\theta)$
and sampled $\mathbf{\mathbf{r}}\sim p_{R}(\mathbf{r})$, the reconstructed
sample $\mathbf{H}=g(\mathbf{z},\mathbf{r};\theta)$, where $g=f^{-1}$,
has a marginal distribution $q(\mathbf{H})$ that matches the true
distribution $p_{H}(\mathbf{H})$, i.e.,
\begin{equation}
\mathrm{MMD}^{2}(q(\mathbf{H}),p_{H}(\mathbf{H}))=0.
\end{equation}
\end{thm}
The proof is given in Appendix \ref{sec:Proof-of-thm:pdf_match}.

\subsubsection{Backward Loss}

For the backward pass, we penalize the reconstruction error between
the backward output $f^{-1}(\mathbf{z},\mathbf{r};\theta)$ and the
original CSI $\mathbf{H}$ with a loss $\mathcal{L}_{\mathbf{H}},$
given by
\begin{equation}
\mathcal{L}_{\mathbf{H}}(g(\mathbf{z},\mathbf{r};\theta),\mathbf{H})=\mathbb{E}\left[\left\Vert \mathbf{H}-g(\mathbf{z},\mathbf{r};\theta)\right\Vert _{F}^{2}\right],\label{eq:loss_mse}
\end{equation}
where $\mathcal{L}_{\mathbf{H}}$ is implemented through Mean Square
Error (MSE).

Although $\mathcal{L}_{\mathbf{r}}$ is theoretically sufficient for
distribution alignment as shown in Theorem \ref{thm:pdf_match}, the
guarantee of distributional matching depends on the condition that
$\mathcal{L}_{\mathbf{r}}$ goes to zero, which is hard to achieve
during the training process. As a result, residual dependencies between
$\mathbf{z}$ and $\mathbf{r}$ may still exist. Moreover, $\mathcal{L}_{\mathbf{r}}$
ensures only distribution-level alignment, and cannot guarantee sample-wise
accuracy. To address these issues, we further impose loss $\mathcal{L}_{\mathbf{H}}$
in the input domain to directly penalize the mismatch between the
constructed CSI and the original input, thereby reinforcing data alignment
and improving the performance in practical training.

\subsubsection{MMD Calculation of $\mathcal{L}_{\mathbf{r}}$}

Let $\mathbf{x}=[\mathbf{z},\mathbf{r}]$ be the samples from the
forward output and let $\mathbf{x}'=[\mathbf{z},\mathbf{\mathbf{\mathbf{r}}}]$
be the samples drawn from $p(\mathbf{z})p_{R}(\mathbf{\mathbf{\mathbf{r}}})$.
The empirical MMD between these two distributions can be computed
using a kernel $k(\cdot,\cdot)$ as 
\begin{align}
\mathrm{MMD}^{2}\left(q(\mathbf{z},\mathbf{r}),p(\mathbf{z})p_{R}(\mathbf{r})\right)=\frac{1}{L^{2}}\sum_{i,j}k_{\text{joint}}\left(\mathbf{x}_{i},\mathbf{x}_{j}\right)\nonumber \\
+\frac{1}{L^{2}}\sum_{i,j}k_{\text{joint}}\left(\mathbf{x}'_{i},\mathbf{x}'_{j}\right)-\frac{2}{L^{2}}\sum_{i}\sum_{j}k_{\text{joint}}\left(\mathbf{x}_{i},\mathbf{x}'_{j}\right),\label{eq:MMD}
\end{align}
where
\begin{equation}
k_{\text{joint}}\left(\mathbf{x}_{i},\mathbf{x}_{j}\right)=k(\mathbf{z}_{i},\mathbf{z}_{j})k(\mathbf{r}_{i},\mathbf{r}_{j}),
\end{equation}
$L$ stands for the number of samples. Here we consider the Inverse
Multiquadric Kernel \cite{tolstikhin2017wasserstein}, that is: 
\begin{equation}
k(\mathbf{x},\mathbf{y})=\frac{C}{C+\Vert\mathbf{x}-\mathbf{y}\Vert_{2}^{2}}.\label{eq:kernel}
\end{equation}

Hence, the training objective of the entire INN-based CSI feedback
system is the weighted sum of equations \eqref{eq:loss_DistributionMatch}
and \eqref{eq:loss_mse}, i.e.,: 
\begin{equation}
\mathcal{L}=\mathcal{L}_{\mathbf{H}}+\kappa\cdot\mathcal{L}_{\mathbf{r}}\label{eq:combined_loss}
\end{equation}
where $\kappa>0$ is the trade-off parameters.

\section{InvCSINet: Robust CSI Feedback with Quantization and Channel Errors\label{sec:InvCSINet:-Robust-CSI}}

In this section, we extent the INN-based CSI feedback framework to
address practical impairments caused by quantization and wireless
channel errors. These practical operations present two primary challenges:
1) quantization and bit-channel distortion are inherently non-differentiable,
making them incompatible with gradient-based training; 2) both operations
introduce irreversible information loss, potentially degrading the
CSI recovery accuracy. To address these issues, we propose InvCSINet,
a robust, end-to-end trainable framework that incorporates information
compensation mechanisms to preserve the integrity of the feedback
process.

\subsection{Overall Framework of InvCSINet}

\begin{figure*}
\centering{}\includegraphics[viewport=0bp 80bp 842bp 510bp,clip,scale=0.62]{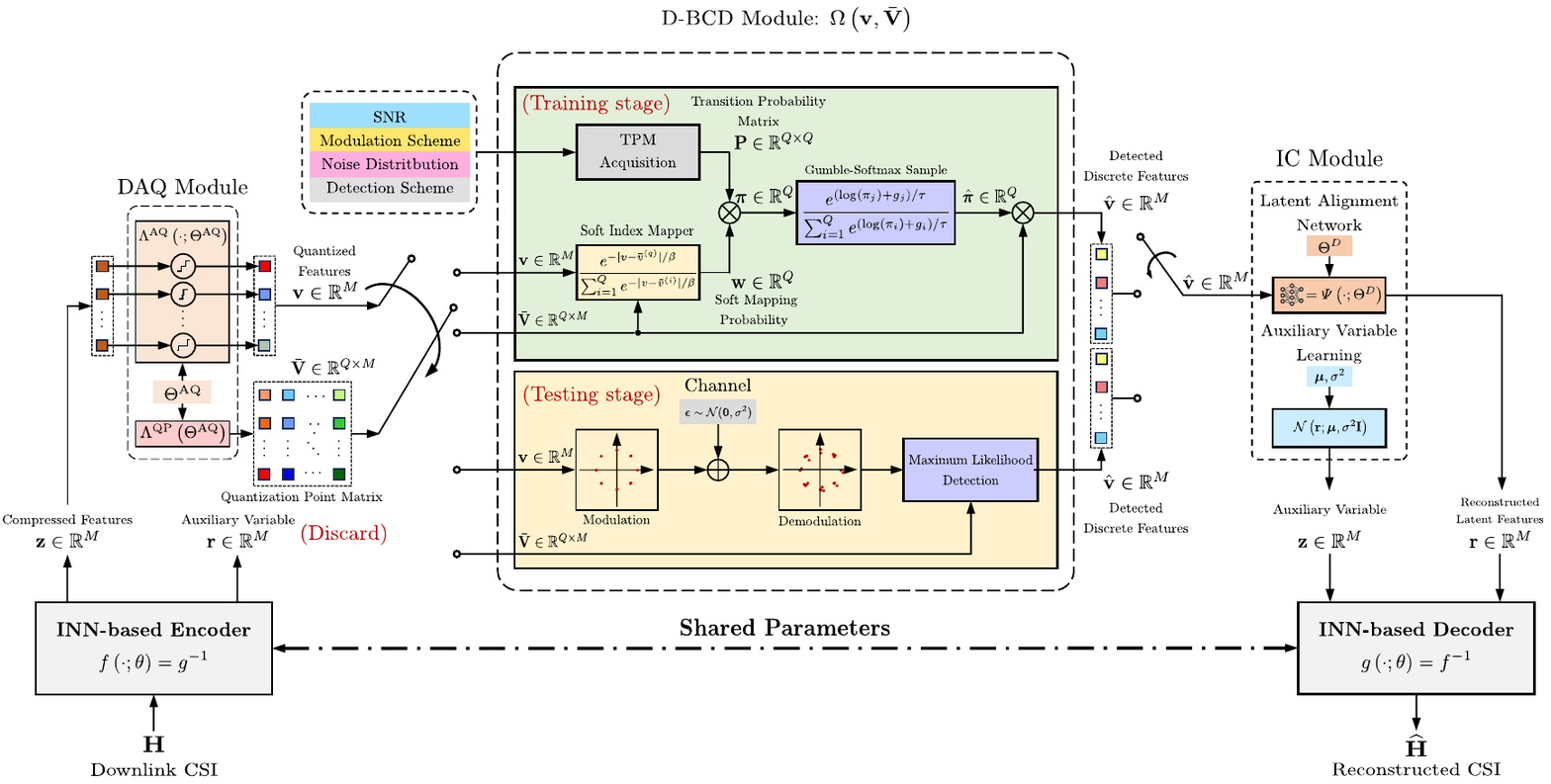}\caption{Network framework of InvCSINet.\label{InvCSINet}}
\end{figure*}

The overall framework of InvCSINet is illustrated in Fig. \ref{InvCSINet}.
The core components are outlined below:
\begin{enumerate}
\item INN-based Encoder and Decoder: As in the ideal transmission scenario,
CSI $\mathbf{H}$ is first preprocessed through domain transformation,
after which it is segmented into two parts, $\mathbf{H}_{1}$ and
$\mathbf{H}_{2}$. These two segments are then input into the INN
to generate a compressed feature $\mathbf{z}$ and an auxiliary feature
$\mathbf{r}$. At the BS, the reconstructed feature $\hat{\mathbf{z}}$
and a sampled auxiliary variable $\mathbf{r}$ are fed into the inverse
path of the INN to reconstruct the original CSI $\mathbf{H}$.
\item DAQ Module: This module approximates quantization in a differentiable
manner, allowing it to be integrated into the back-propagation training.
It also supports an adaptive non-uniform quantization scheme that
dynamically adjusts quantization points across feature dimensions
based on their individual distributions. This enables the network
to lower the quantization distortion under constrained feedback rates
and enhances compression efficiency. The details will be elaborated
in Section \ref{subsec:Adaptive-Quantization-Module}.
\item D-BCD Module: Built upon the Gumbel-Softmax technique, this module
simulates the bit transmission process and accounts for potential
errors introduced by channel noise. By modeling the digital transmission
process in a differentiable manner, this module enables end-to-end
training and improves the network's robustness to channel impairments,
which will be introduced in detail in Section \ref{subsec:Differentiable-Bit-Channel-Disto}.
\item IC Module : To mitigate the distortion caused by quantization and
channel errors and thereby preserve information throughout the process,
a LAN and a learnable auxiliary variable are incorporated. The LAN
corrects reconstruction errors, thereby aligning the distribution
of $\hat{\mathbf{z}}$ with that of $\mathbf{z}$. The auxiliary variable
captures residual information lost during transmission. Further details
are provided in Section\,\ref{subsec:Information-Compensation-Module}.
\end{enumerate}

\subsection{Differentiable Adaptive Quantization Module\label{subsec:Adaptive-Quantization-Module}}

The DAQ module takes the compressed feature $\mathbf{z}\in\mathbb{R}^{M}$
as input and produces a discretized representation $\mathbf{v}\in\mathbb{R}^{M}$.
Assume each element of $\mathbf{z}$ is quantized into $B$ bits.
The entire DAQ process can be represented as
\begin{align}
\mathbf{v} & =\Lambda^{\mathrm{DAQ}}(\mathbf{z};\Theta^{\mathrm{DAQ}});\bar{\mathbf{V}}=\Lambda^{\mathrm{QP}}(\Theta^{\mathrm{DAQ}}),
\end{align}
where $\mathbf{v}$ denotes the quantized feature vector and $\bar{\mathbf{V}}=\left[\bar{\mathbf{v}}_{1},\cdots,\bar{\mathbf{v}}_{M}\right]\in\mathbb{\mathbb{R}}^{Q\times M}$
stands for the quantization point matrix, with $Q=2^{B}$ denoting
the number of quantization levels. Each column $\bar{\mathbf{v}}_{i}=\left[\bar{v}_{i}^{(1)},\cdots,\bar{v}_{i}^{(Q)}\right]^{T}$
corresponds to the collection of quantization points for the $i$-th
dimension of $\mathbf{z}$. $\Theta^{\mathrm{DAQ}}$ denotes the set
of learnable parameters involved in the DAQ process. The detailed
design of the DAQ module will be elaborated in the following.

Denote the $i$-th element of $\mathbf{z}$ as $z_{i}$. We firstly
consider an ideal uniform quantization scheme, which uniformly divides
the range of $z_{i}$ into $Q$ quantization intervals. The quantization
resolution is defined as $r_{i}=\frac{z_{max}-z_{min}}{Q}$, where
$(z_{min},z_{max})$ represents the predefined range. Then, the ideal
uniform quantization function can be expressed as
\begin{equation}
\Lambda(z_{i})=\frac{z_{max}+z_{min}}{2}+\frac{r_{i}}{2}\sum_{q=1}^{Q-1}\epsilon\left(z_{i}-(z_{min}+qr_{i})\right),
\end{equation}
where $\epsilon(\cdot)$ is a discontinuous sign function, given by
\begin{equation}
{\epsilon(x)}=\begin{cases}
1, & {\text{if}}\ x>0\\
0, & {\text{if}}\ x=0\\
-1, & {\text{if}}\ x<0
\end{cases}.
\end{equation}
Due to the discontinuity of $\epsilon(\cdot)$, the ideal quantization
function is non-differentiable. To ensure differentiability and reduce
quantization distortion, the proposed DAQ module models the quantization
process as a linear combination of differentiable functions with learnable
biases and scales, which consists of the following two operations.

\subsubsection{Soft Quantization Function}

We approximate the sign function $\epsilon(x)$ using a continuous
and smooth $S$-shaped function \cite{cao2017hashnet}:
\begin{equation}
\hat{\epsilon}(x)=\frac{Tx}{1+\vert Tx\vert},\label{eq:S-approx}
\end{equation}
where $T$ controls the degree of approximation, with larger $T$
yielding a closer approximation to the ideal function. The soft quantization
function enables the network to account for quantization effects during
training without significant gradient mismatch between training and
inference.

\subsubsection{AQ Function}

The AQ function introduces learnable bias and scaling factors. For
the $i$-th element, the quantized feature can be written as
\begin{equation}
v_{i}=\Lambda^{\mathrm{DAQ}}(z_{i})=c_{i}+\sum_{q=1}^{Q-1}a_{q,i}\hat{\epsilon}\left(z_{i}-b_{q,i}\right),
\end{equation}
where the learnable parameters $\text{\ensuremath{\Theta}}^{\mathrm{DAQ}}=\left\{ a_{q,i},b_{q,i},c_{i}\right\} _{q=1,i=1}^{Q-1,M}$
should satisfy the following conditions:
\begin{equation}
b_{1,i}\leq b_{2,i}\leq\cdots\leq b_{Q-1,i},
\end{equation}
\begin{equation}
a_{q,i}\geq0,\quad\forall q,i.
\end{equation}
Note that the translation bias $\{b_{q,i}\}$ define the boundaries
of the quantization regions. The scaling factors $\{a_{q,i}\}$ specify
the step sizes between adjacent quantization levels. The pairs $\{a_{q,i},c_{i}\}$
jointly define the quantization levels $\{\bar{v}_{i}^{(q)}\}$ within
each region. Specifically, the relationship between the DAQ parameters
$\Theta^{\mathrm{DAQ}}$ and the quantization point matrix $\bar{\mathbf{V}}$
can be characterized by the mapping function $\Lambda^{\mathrm{QP}}$,
which determines the quantization points for each feature dimension
through the following equations:
\begin{align}
\bar{v}_{i}^{(1)} & =c_{i}-\sum_{q=1}^{Q-1}a_{q,i},.\\
\bar{v}_{i}^{(q)} & =\bar{v}_{i}^{(q-1)}+2a_{q-1,i},2\leq q\leq Q-1.
\end{align}

Due to the differentiable AQ function, the quantization parameters
$\{a_{q,i},b_{q,i},c_{i}\}$ can be learned end-to-end by minimizing
the total loss function, enabling a distortion-aware, non-uniform
quantization scheme that minimizes the quantization error under a
given feedback constraint.

\subsection{Differentiable Bit-Channel Distortion Module\label{subsec:Differentiable-Bit-Channel-Disto}}

Since the quantized feature vector $\mathbf{v}$ is transmitted sequentially
on a per-dimension basis, we model the bit-level transmission and
channel distortion independently for each dimension. For clarity,
we omit the feature dimension index in this subsection. Let $v$ denote
the quantization output at the user. Let $\hat{v}$ denote the detection
output at the BS. The D-BCD module can be characterized by a differentiable
function $\Omega$ as:
\begin{equation}
\hat{v}=\Omega(v,\bar{\mathbf{v}}),
\end{equation}
where $\bar{\mathbf{v}}=\left[\bar{v}^{(1)},\dots,\bar{v}^{(Q)}\right]^{T}$
is the quantization point vector for $v$ in the DAQ module. The D-BCD
module comprises several operations, as illustrated below.

\subsubsection{Soft Index Mapper Output ($\mathbf{w}$)}

Due to soft quantization, the output $v$ may not exactly lie on the
quantization points. To quantify this deviation, we introduce a soft
mapping probability vector $\mathbf{w}\in[0,1]^{Q\times1}$, where
each entry $w_{q}$ represents the likelihood that $v$ corresponds
to the $q$-th quantization point $\bar{v}^{(q)}$. Specifically,
$w_{q}$ is defined as
\begin{equation}
w_{q}=\frac{\exp\left(-\frac{\left|v-\bar{v}^{(q)}\right|}{\beta}\right)}{\sum_{i=1}^{Q}\exp\left(-\frac{\left|v-\bar{v}^{(i)}\right|}{\beta}\right)},
\end{equation}
where $\beta$ is a small positive constant that controls the sharpness
of the distribution. When an ideal quantization function is applied
during the inference, the vector $\mathbf{w}$ becomes a one-hot vector,
indicating a hard assignment to the nearest quantization point.

\subsubsection{Transition Probability Matrix Acquisiton ($\mathbf{P}$)}

Assume the $q$-th quantization point is selected for transmission,
i.e, $v=\bar{v}^{(q)}$. This symbol undergoes binary encoding followed
by digital modulation, after which the modulated signal is transmitted
over a noisy communication channel. At the BS, the received noisy
symbol is processed via maximum likelihood (ML) detection to determine
the most likely transmitted symbol. The detected symbol is then demodulated
and decoded to recover the corresponding quantized representation,
denoted as $\hat{v}$. We model this entire transmission process using
a transition probability matrix (TPM) $\mathbf{P}\in[0,1]^{Q\times Q}$,
where the $(i,j)$-th element corresponds to the probability that
the $j$-th quantization symbol $\bar{v}^{(j)}$ is transmitted, and
the $i$-th quantization symbol $\bar{v}^{(i)}$ is detected at the
receiver, i.e., 
\begin{equation}
p_{ij}=P\left(\hat{v}=\bar{v}^{(i)}|v=\bar{v}^{(j)}\right).
\end{equation}
Based on the ML detection principle, $p_{ij}$ equals to the probability
that the received noisy signal falls within the decision region associated
with $\bar{v}^{(i)}$, i.e., 
\begin{equation}
p_{ij}=P\left(i=\arg\max_{q\in\{1,\dots,Q\}}p(\hat{v}=\bar{v}^{(q)}|v=\bar{v}^{(j)})\right).
\end{equation}
The TPM $\mathbf{P}$ depends on several factors, including the signal-to-noise
ratio (SNR) of the channel and the modulation scheme employed. For
example, consider a standard Binary Phase Shift Keying (BPSK) modulation
scheme under an additive white Gaussian noise (AWGN) channel with
SNR denoted as $\gamma$. Let the binary representation of $\bar{v}^{(j)}$
be $U_{1}^{(j)}U_{2}^{(j)}\dots U_{B}^{(j)}$, where $B$ is the number
of bits and $U_{k}^{(j)}\in\{0,1\}$. Similarly, denote the binary
representation of $\bar{v}^{(i)}$ as $U_{1}^{(i)}U_{2}^{(i)}\dots U_{B}^{(i)}$.
The transition probability $p_{ij}$ can be calculated as:
\begin{equation}
p_{ij}=\left[Q(\sqrt{\gamma})\right]^{D_{\mathrm{ham}}}\left[1-Q(\sqrt{\gamma})\right]^{B-D_{\mathrm{ham}}},
\end{equation}
where $Q(\cdot)$ is the tail distribution function of the standard
normal distribution, and $D_{\mathrm{ham}}$ is the Hamming distance
between the two binary sequences. Note that this soft transmission
model can be extended to other modulation schemes and fading channel
models by appropriately modifying the computation of the TPM $\mathbf{P}.$

\subsubsection{Gumble-Softmax Sampling}

Based on the probability vector $\mathbf{w}$ and the TPM $\mathbf{P}$,
we can derive the categorical distribution of the detected discrete
representation $\hat{v}$. Specifically, the probability distribution
$\bm{\pi}\in[0,1]^{Q\times1}$ of $\hat{v}$ can be calculated through
\begin{equation}
\bm{\pi}=\mathbf{P}\mathbf{w},
\end{equation}
where the $q$-th entry $\pi_{q}$ denotes the probability that the
$q$-th quantization point is detected at the BS, i.e., $\pi_{q}=P\left(\hat{v}=\bar{v}^{(q)}\right)$.
Directly sampling from the categorical distribution $\bm{\pi}$ is
non-differentiable, posing challenges for gradient-based optimization.
To address this, we adopt the Gumbel-Softmax trick \cite{jang2016categorical}.
This technique enables differentiable approximation of categorical
sampling by perturbing the logits with Gumbel noise and applying the
softmax function. As a result, it facilitates end-to-end training
of neural networks involving discrete variables. Specifically, to
obtain a continuous relaxation of categorical sampling for $\hat{v}$,
we compute a sample vector $\mathbf{\hat{\boldsymbol{\pi}}}\in\mathbb{R}^{Q}$,
whose $i$-th entry is given by
\begin{equation}
\hat{\pi}_{j}=\frac{\exp\left((l_{j}+g_{j})/\tau\right)}{\sum_{i=1}^{\mathop{Q}}\exp\left((l_{i}+g_{i})/\tau\right)},\label{eq:prob}
\end{equation}
where $l_{j}=\log(\pi_{j})$ are logits. $\{g_{1},g_{2},\dots,g_{Q}\}$
are i.i.d. samples from the Gumbel distribution computed by $g_{i}=-\log\left(-\log(u_{i})\right)$
where $u_{i}\sim\mathrm{Uniform}(0,1)$. $\tau\ge0$ is the softmax
temperature. As $\tau$ approaches 0, the sample vector $\mathbf{\hat{\boldsymbol{\pi}}}$
is close to the one-hot like vector and the Gumbel-softmax distribution
approaches the categorical distribution $\bm{\pi}$. In this way,
the detected discrete feature $\hat{v}$ can be obtained through
\begin{equation}
\hat{v}=\bar{\mathbf{v}}^{T}\mathbf{\mathbf{\hat{\boldsymbol{\pi}}}},
\end{equation}
where $\bar{\mathbf{v}}=\left[\bar{v}^{(1)},\dots,\bar{v}^{(Q)}\right]^{T}$.

Figure \ref{InvCSINet} illustrates the signal processing flow of
the proposed D-BCD module. During training, the Gumbel-Softmax technique
is employed to approximate the categorical sampling of the detected
symbol, thereby rendering the entire modulation and detection pipeline
differentiable and compatible with end-to-end optimization. During
testing, standard digital modulation and ML detection are applied
to evaluate the system's practical performance.

\begin{figure*}[t]
\begin{centering}
\includegraphics[viewport=140bp 220bp 700bp 380bp,clip,scale=0.82]{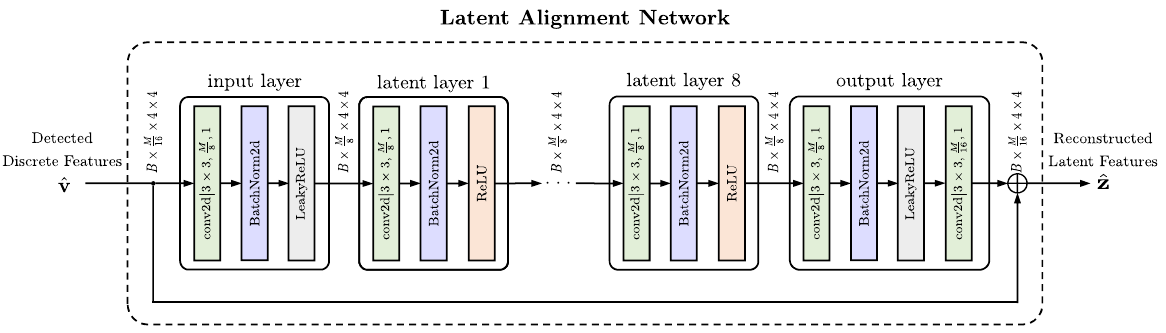}
\par\end{centering}
\caption{The structure of the LAN. Feature size is denoted as: batch size $\times$
channel $\times$ height $\times$ width. \label{fig:compensate}}
\end{figure*}

\subsection{Information Compensation Module \label{subsec:Information-Compensation-Module}}

Due to quantization distortion and channel noise-induced error, the
detected output $\hat{\mathbf{v}}$ suffers from information loss
compared to the original compressed feature $\mathbf{z}$. To compensate
for this information loss, we propose a LAN that takes $\hat{\mathbf{v}}$
as input and outputs the reconstructed latent feature $\hat{\mathbf{z}}$.
The distribution alignment process can be written as 
\begin{equation}
\hat{\mathbf{z}}=\varPsi(\hat{\mathbf{v}};\Theta^{D}),
\end{equation}
where $\Theta^{D}$ stands for the network parameters to be trained.
The structure of the LAN $\varPsi$ is illustrated in Fig. \ref{fig:compensate},
which consists of a residual connection and a CNN. The primary objective
of the denoising network is to mitigate the adverse effects of quantization
and channel noise such that the distribution of the reconstructed
latent features $\mathbf{\hat{z}}$ approximates the marginal distribution
of the compressed features $\mathbf{z}$ at the output of the INN
forward path.

Meanwhile, we propose to learn a prior distribution for the auxiliary
variable $\mathbf{r}$, which is used to sample the auxiliary feature
in the INN inverse process. The learned auxiliary variable can further
capture the information of CSI that is not covered in $\hat{\mathbf{z}}$.
We assume the learned prior follows a Gaussian distribution, i.e.,
$\upsilon_{R}(\mathbf{r})=\mathcal{N}(\boldsymbol{\mu},\sigma^{2}\mathbf{I})$,
where the mean vector $\boldsymbol{\mu}$ and variance $\sigma^{2}$
are treated as trainable parameters during the network optimization.
Specifically, we re-parameterize $\mathbf{r}$ as $\mathbf{r}=\sigma\cdot\mathbf{e}+\bm{\mu}$,
where $\mathbf{e}\sim\mathcal{N}(\bm{0},\mathbf{I})$, allowing the
sampling process to retain gradients. The auxiliary feature $\mathbf{r}$
sampled from the learned distribution $\upsilon_{R}(\mathbf{r}),$
together with the recovered latent feature $\hat{\mathbf{z}}$, is
fed into the inverse process of the INN to recover the original CSI.

\subsection{Overall Training of the InvCSINet}

Based on the INN principle in Theorem \ref{thm:pdf_match}, to guarantee
a distributional fidelity of CSI recovery, in the feature subspcace,
we require that the joint distribution of the compressed feature and
the auxiliary variable at the output of the INN forward path, i.e.,
$q(\mathbf{z},\mathbf{r})$, should match the product of the marginal
distribution of the recovered latent feature and the prior distribution
of the auxiliary variable at the input of the INN inverse path, i.e.,
$p(\hat{\mathbf{z}})\upsilon_{R}(\mathbf{r})$. To enforce this constraint,
the following distribution matching loss function should be minimized:
\begin{equation}
\mathcal{L}_{\mathbf{r}}\left(q(\mathbf{z},\mathbf{r}),p(\hat{\mathbf{z}})\upsilon_{R}(\mathbf{r})\right)=\mathrm{MMD}^{2}\left(q(\mathbf{z},\mathbf{r}),p(\hat{\mathbf{z}})\upsilon_{R}(\mathbf{r})\right),\label{eq:MMD_practical}
\end{equation}
where MMD can be calculated through \eqref{eq:MMD}. We have the following
theorem for the performance of the proposed InvCSINet.
\begin{thm}
[Distributional Fidelity of InvCSINet] \label{thm:pdf_match-1}Let
$p_{H}(\mathbf{H})$ denote the prior distribution of the CSI. Suppose
an INN $f(\mathbf{H};\theta)=[\mathbf{z},\mathbf{r}]$, an DAQ module
$\mathbf{v}=\Lambda^{\mathrm{DAQ}}(\mathbf{z};\Theta^{\mathrm{DAQ}})$
and $\bar{\mathbf{V}}=\Lambda^{\mathrm{QP}}(\Theta^{\mathrm{DAQ}}),$
a D-BCD module $\hat{\mathbf{v}}=\Omega(\mathbf{v},\bar{\mathbf{V}})$,
a denoising network $\hat{\mathbf{z}}=\varPsi(\hat{\mathbf{v}};\Theta^{D})$,
and a learnable auxiliary variable $\upsilon_{R}(\mathbf{r})$ are
jointed trained to minimize the unsupervised loss $\mathcal{L}_{\mathbf{r}}=\mathrm{MMD}^{2}(q(\mathbf{z},\mathbf{r}),p(\hat{\mathbf{z}})\upsilon_{R}(\mathbf{r}))$,
where $\upsilon_{R}(\mathbf{r})=\mathcal{N}(\mathbf{r};\boldsymbol{\mu},\sigma^{2}\mathbf{I})$,
and assume the loss reaches zero. Then, for any input $\hat{\mathbf{z}}=\varPsi\left(\Omega\left(\Lambda^{\mathrm{DAQ}}\left(f_{\mathbf{z}}(\mathbf{H};\theta)\right),\bar{\mathbf{V}}\right)\right)$
and sampled $\mathbf{r}\sim\upsilon_{R}(\mathbf{r})$, the reconstructed
sample $\mathbf{H}=g(\hat{\mathbf{z}},\mathbf{r};\theta)$, where
$g=f^{-1}$, has a marginal distribution $q(\mathbf{H})$ that matches
the true distribution $p_{H}(\mathbf{H}$), i.e.,
\begin{equation}
\mathrm{MMD}^{2}\left(q(\mathbf{H}),p_{H}(\mathbf{H})\right)=0.
\end{equation}
\end{thm}
The proof is given in Appendix \ref{sec:Proof-of-thm:pdf_match-1}.
Theorem \ref{thm:pdf_match-1} shows that when the network converges
such that the distribution matching loss $\mathcal{L}_{\mathbf{r}}$
in \eqref{eq:MMD_practical} approaches zero, the recovered latent
feature $\hat{\mathbf{z}}$ at the BS becomes equivalent to the compressed
feature $\mathbf{z}$ at the user side. Consequently, according to
Theorem \ref{thm:pdf_match}, the distribution of the recovered CSI
obtained through InvCSINet converges to the prior distribution of
the original CSI. However, in practical training, it is often challenging
to minimize the MMD in $\mathcal{L}_{\mathbf{r}}$ to zero, especially
with the information loss introduced by quantization and channel noise.
To enhance recovery performance under such conditions, we introduce
an additional reconstruction loss $\text{\ensuremath{\mathcal{L}}}_{\mathbf{H}}$
to penalize the mismatch between the original CSI and the final reconstructed
CSI to further regulate the whole CSI feedback pipeline. Therefore,
the overall training loss function for the proposed InvCSINet is given
by 
\begin{equation}
\mathcal{L}_{\text{total}}=\mathcal{L}_{\mathbf{H}}+\kappa\cdot\mathcal{L}_{\mathbf{r}},
\end{equation}
where $\mathcal{L}_{\mathbf{H}}(g(\hat{\mathbf{z}},\mathbf{\mathbf{r}};\theta),\mathbf{H})=\mathbb{E}\left[\left\Vert \mathbf{H}-g(\hat{\mathbf{z}},\mathbf{\mathbf{r}};\theta)\right\Vert _{F}^{2}\right]$
and $\mathcal{L}_{\mathbf{r}}$ is given by \eqref{eq:MMD_practical}
.

\section{Simulation Results\label{sec_4} }

This section evaluates the performance of InvCSINet in FDD massive
MIMO systems. The Normalized Mean Square Error (NMSE) is primarily
used as the performance metric for final channel recovery, defined
as
\begin{equation}
\text{NMSE}(\widehat{\mathbf{H}},\mathbf{H})=\mathbb{E}\left[\frac{\Vert\widehat{\mathbf{H}}-\mathbf{H}\Vert_{F}^{2}}{\Vert\mathbf{H}\Vert_{F}^{2}}\right].
\end{equation}

\subsection{Simulation Setups}

\subsubsection{Datasets}

We adopt DeepMIMO \cite{alkhateeb2019deepmimo} to generate the dataset.
Consider an outdoor scene with two streets and an intersection, operating
at 28 GHz and with a bandwidth of 50 MHz. The number of subcarriers
is $N_{c}=1024$. The BS has $N_{t}=32$ antennas and the user has
$N_{r}=32$ antennas. The generated channel data have been normalized
and standardized to reduce the adverse effects of singular sample
data.

\subsubsection{Network Implementation}

\begin{figure}
\subfloat[The structure of $\bm{\phi}$.\label{fig:The-structure-of-phi}]{\begin{centering}
\includegraphics[viewport=275bp 300bp 550bp 460bp,clip,scale=0.42]{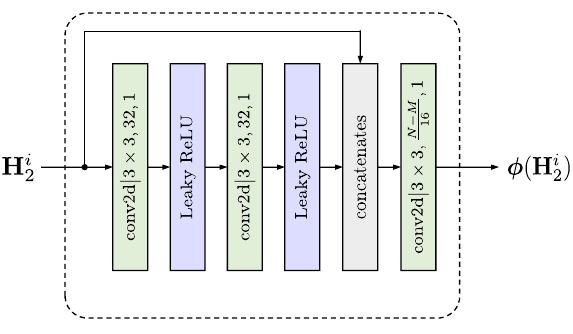}
\par\end{centering}
}\subfloat[The structure of $\bm{\eta}$.\label{fig:The-structure-of-eta}]{\begin{centering}
\includegraphics[viewport=265bp 136bp 560bp 296bp,clip,scale=0.42]{img/phi_eta}
\par\end{centering}
}

\caption{The structure of $\bm{\phi}$ and $\bm{\eta}$.\textcolor{blue}{{} }Each
convolutional layer is denoted as: kernel size, number of output channels,
and padding size. The Concatenate layer merges two feature maps along
the channel dimension. \label{fig:The-structure-of-phi-eta}}
\end{figure}

In the INN architecture, the structures of components $\bm{\phi}$
and $\bm{\eta}$ are illustrated in Fig. \ref{fig:The-structure-of-phi}
and Fig. \ref{fig:The-structure-of-eta}, respectively. The Residual
structure enables intermediate outputs from certain layers to be quickly
fed back into earlier layers, which helps mitigate the vanishing gradient
problem while maintaining strong performance. To simplify the network
design, we set $\boldsymbol{\rho}$ to be an identity operator. The
number of invertible blocks is set to 3. In the DAQ module, the parameters
$\text{\ensuremath{\Theta}}^{\mathrm{DAQ}}=\left\{ a_{q,i},b_{q,i},c_{i}\right\} _{q=1,i=1}^{Q-1,M}$
are initialized to the corresponding values of uniform quantization.
The predefined range $(z_{min},z_{max})$ is initialized as $(-2,2)$.
In the D-BCD module, we adopt a standard BPSK modulation scheme and
an AWGN feedback channel. Additionally, we employ $\texttt{one\_hot}\left(\arg\max_{j}\hat{\pi}_{j}\right)$
to replace the continuous relaxation version of $\hat{\bm{\pi}}$
while retaining the original gradient. This ensures that the decoder
handles discretized features consistently during both training and
testing, preventing distribution mismatches. The structure of IC Module
is shown in Fig. \ref{fig:compensate}. A Flatten layer and an Unflatten
layer are incorporated before the DAQ module and after the IC module
to reshape the dimensions of compressed feature $\mathbf{z}$ and
auxiliary feature $\mathbf{r}$, whose values are determined by the
actual compression ratio. Both the weights and biases of the IC module
are initialized to 0.

\subsubsection{Baselines}

We compare the proposed InvCSINet with four baseline schemes to benchmark
its performance. 
\begin{itemize}
\item \textbf{CsiNet}: The CsiNet \cite{wen2018deep} is a CNN-based DAE
framework trained with MSE loss, assuming ideal transmission without
quantization and noisy feedback channel.
\item \textbf{CsiNet+}: The CsiNet+ \cite{chen2019novel} enhances CsiNet
by introducing a $\mu$-law quantizer and an offline dequantizer trained
via a three-step training strategy minimizing MSE and a regularization
term. The CsiNet+ does not support end-to-end training due to its
non-differentiable quantizer and assumes perfect feedback transmission.
\item \textbf{DNNet}: The DNNet \cite{ye2020deep} uses a DAE structure
with a noise extraction unit (NEU) to mitigate AWGN feedback channel
noise, but does not consider quantization. A two-step training strategy
is employed to alternately minimize the MSE of CSI reconstruction
and NEU for denoising.
\item \textbf{ATNet}: The ATNet \cite{chang2021deep} employs an attention-based
encoder-decoder with an error correction block (ECBlock) to handle
quantization and channel noise. Similar to DNNet, a two-step training
strategy is employed to alternately minimize the MSE of CSI reconstruction
and ECBlock for error correction.
\end{itemize}
In our experiments, all of the above baselines are trained following
the procedures in their respective reference papers, and are tested
in realistic system settings, considering the full transmission pipeline
including quantization, modulation, channel noise, demodulation, and
detection.

\subsubsection{Training Details}

The entire data set is randomly split, with 70\% serving as the training
set and 30\% serving as the testing set. The training dataset contains
12,670 samples, with a batch size of 128, and the testing dataset
comprises 5,430 samples. The training epoch is 1,000. Gradient back-propagation
is performed using the Adam optimizer with default settings \cite{KingBa15}.
The initial learning rate is $10^{-3}$ and will be reduced to 90\%
of the previous value every 20 epochs. The hyper-parameter $\kappa$
is set to 0.1 and the constant $C$ is set to be $1000$ in the MMD
kernel function \eqref{eq:kernel} throughout the experiments. The
experiments were conducted on a server equipped with a Xeon Gold 6354
CPU, 512 GB of RAM, and four NVIDIA RTX 3090 GPUs. The entire system
was implemented using the PyTorch library with Python version 1.10.2.

\begin{figure}
\subfloat[\label{fig:NMSE-versus-the}]{\centering{}\includegraphics[scale=0.52]{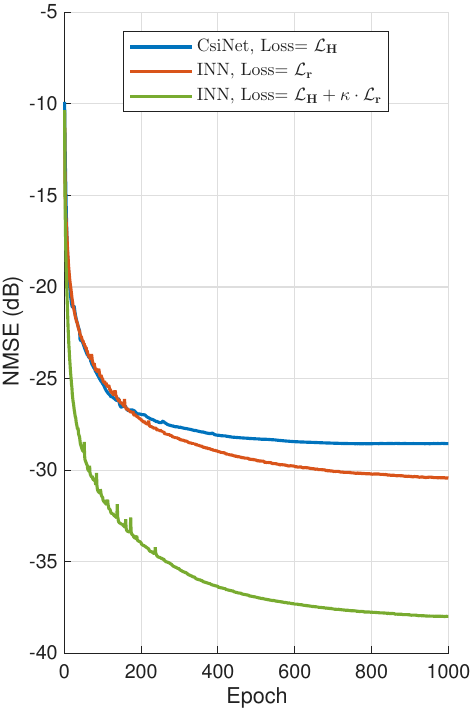}}\subfloat[\label{fig:MMD-versus-the}]{\begin{centering}
\includegraphics[scale=0.52]{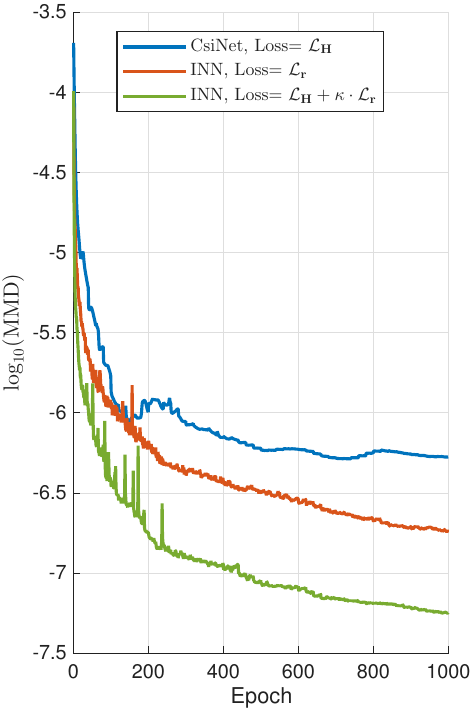}
\par\end{centering}
}

\caption{(a): NMSE performance versus the training epochs at a compression
ratio of 1/32; (b): MMD versus the training epochs at a compression
ratio of 1/32.}
\end{figure}

\begin{figure*}[!t]
\subfloat[]{\begin{centering}
\includegraphics[width=0.23\textwidth]{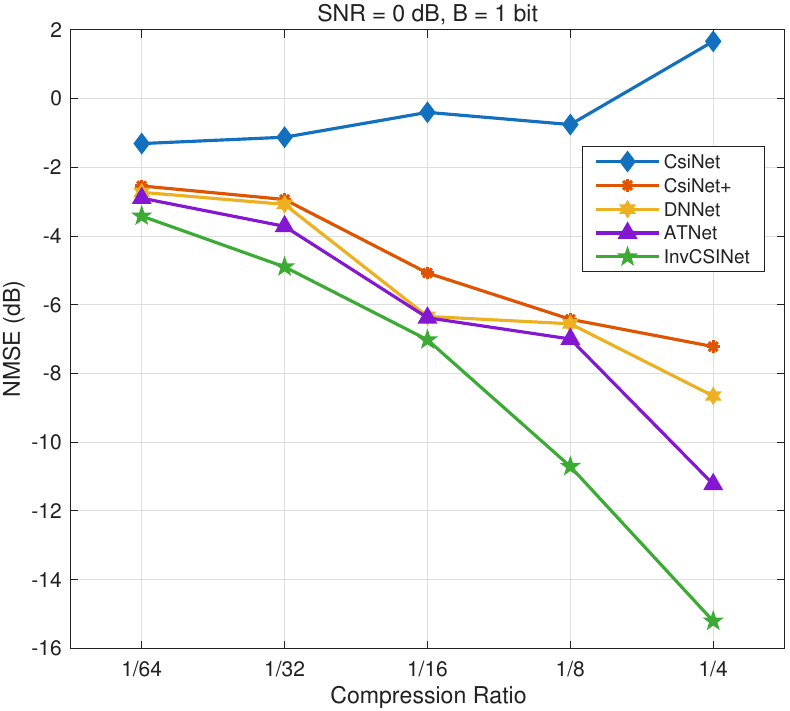}
\par\end{centering}
}\subfloat[]{\begin{centering}
\includegraphics[width=0.23\linewidth]{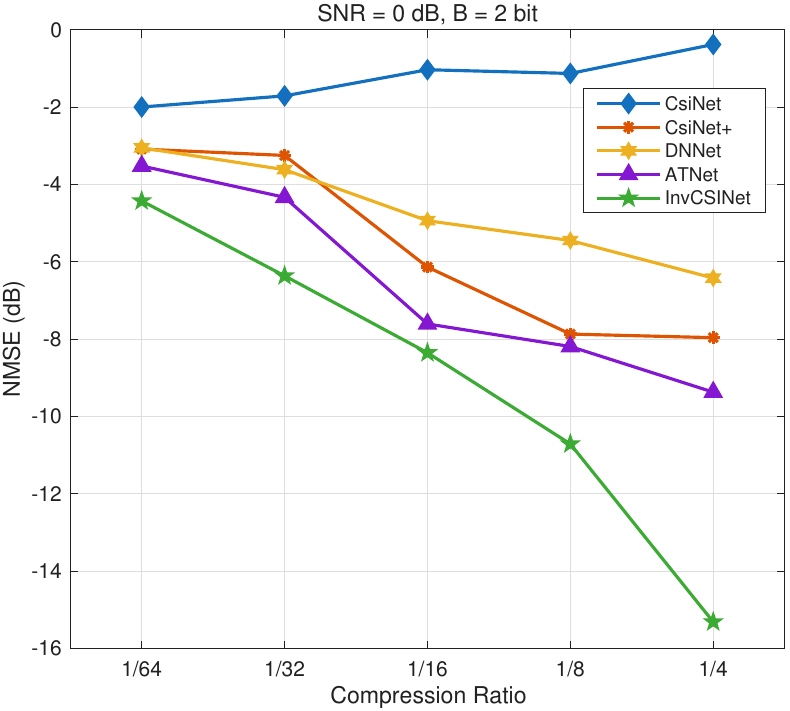}
\par\end{centering}
}\subfloat[]{\begin{centering}
\includegraphics[width=0.23\linewidth]{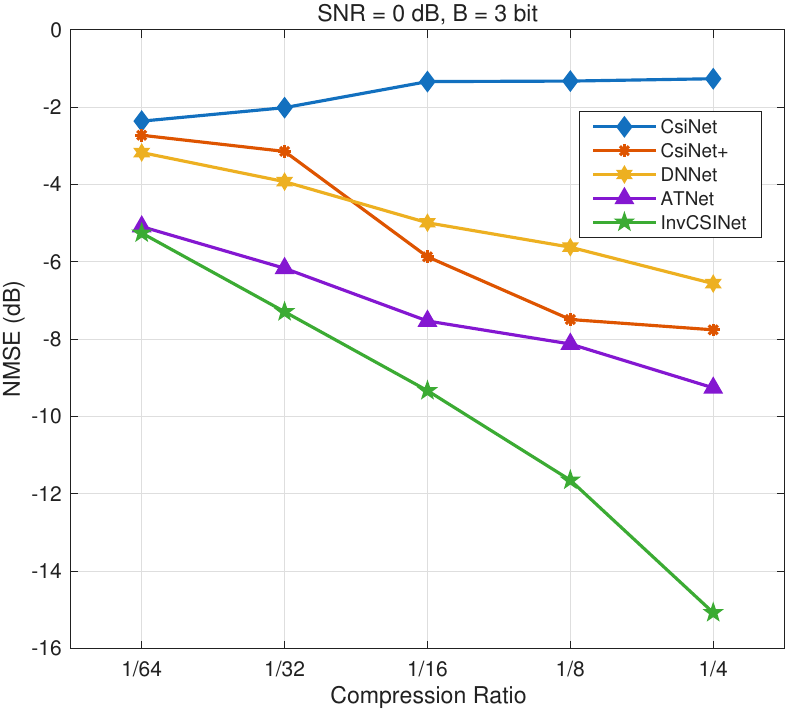}
\par\end{centering}
}\subfloat[]{\begin{centering}
\includegraphics[width=0.23\linewidth]{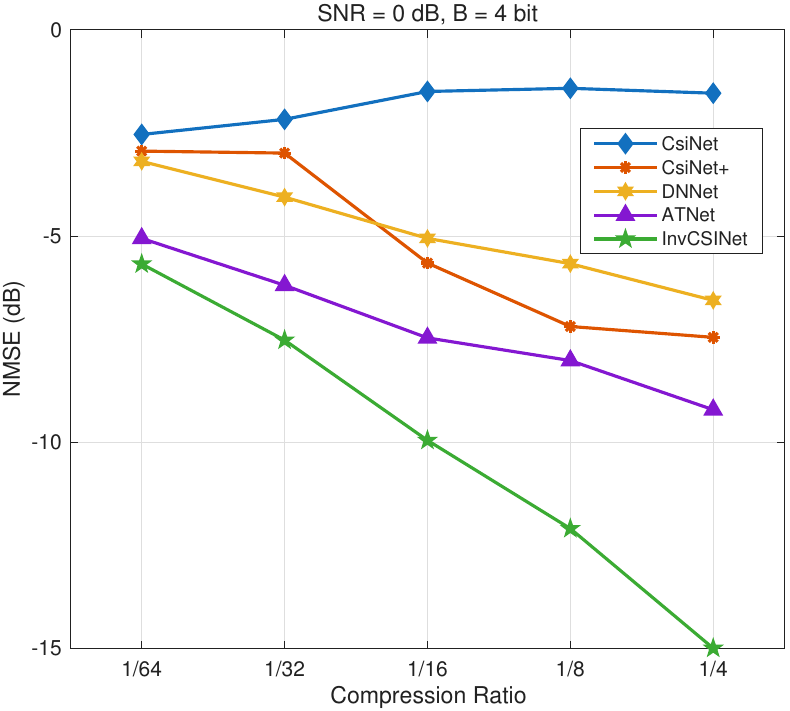}
\par\end{centering}
}

\subfloat[]{\begin{centering}
\includegraphics[width=0.23\linewidth]{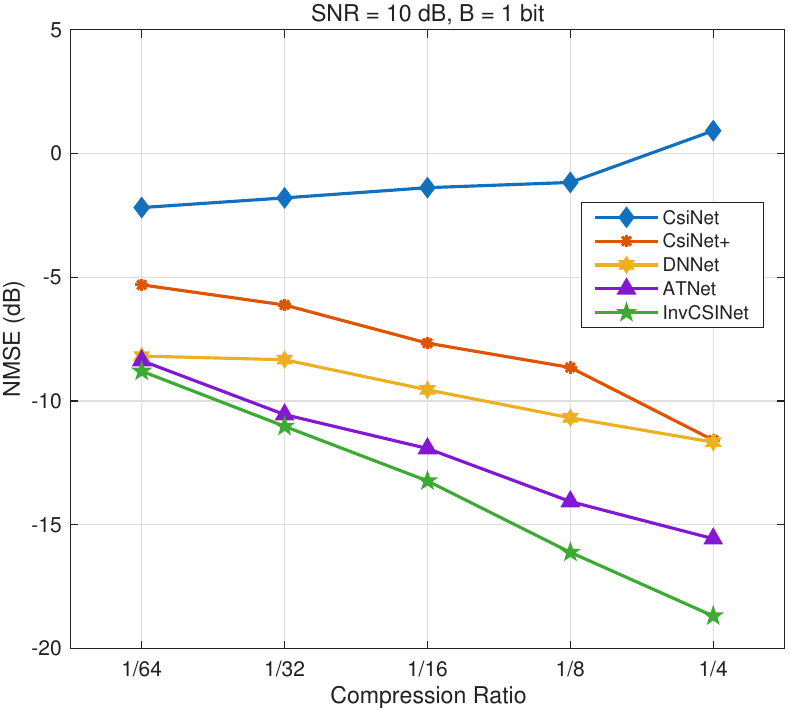}
\par\end{centering}
}\subfloat[]{\begin{centering}
\includegraphics[width=0.23\linewidth]{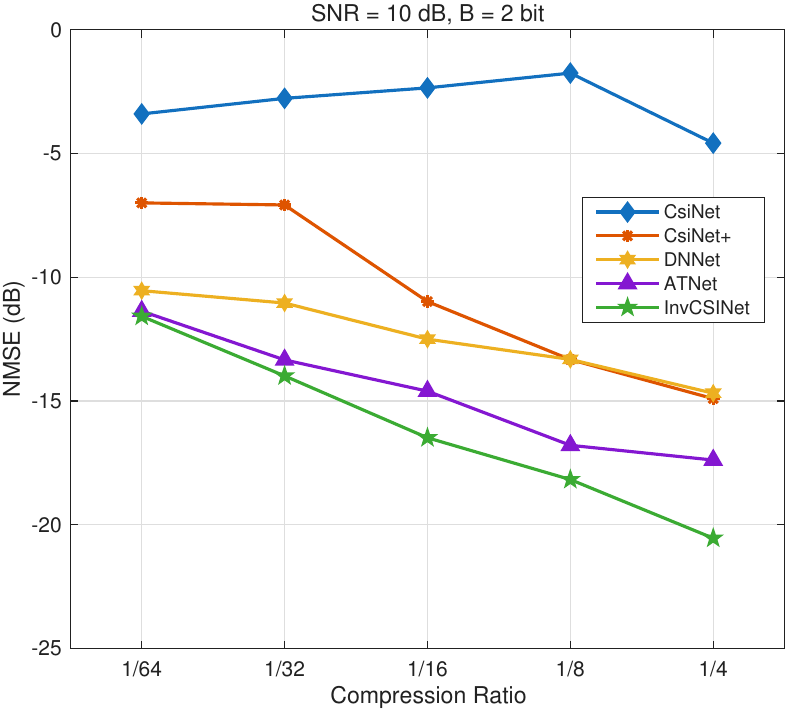}
\par\end{centering}
}\subfloat[]{\begin{centering}
\includegraphics[width=0.23\linewidth]{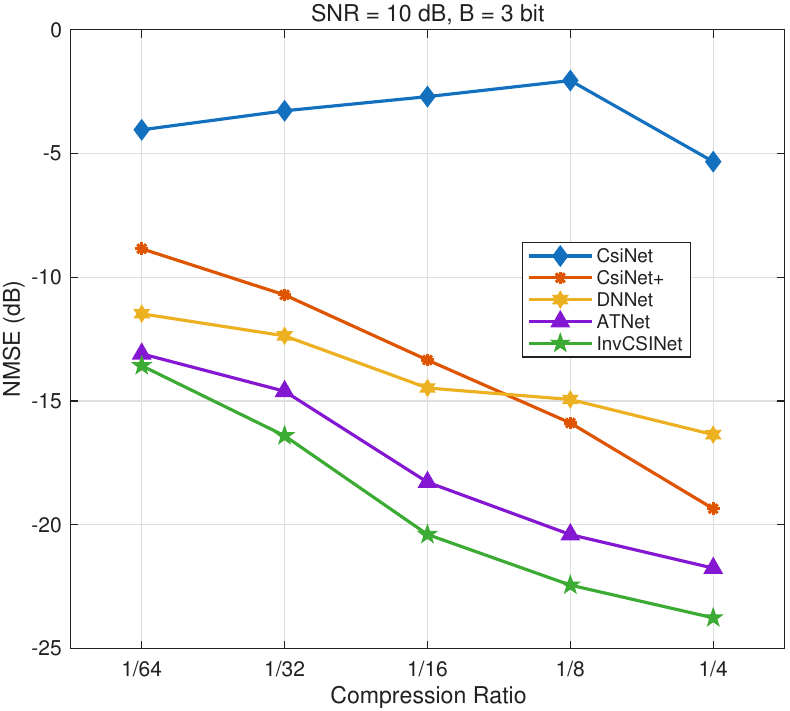}
\par\end{centering}
}\subfloat[]{\begin{centering}
\includegraphics[width=0.23\linewidth]{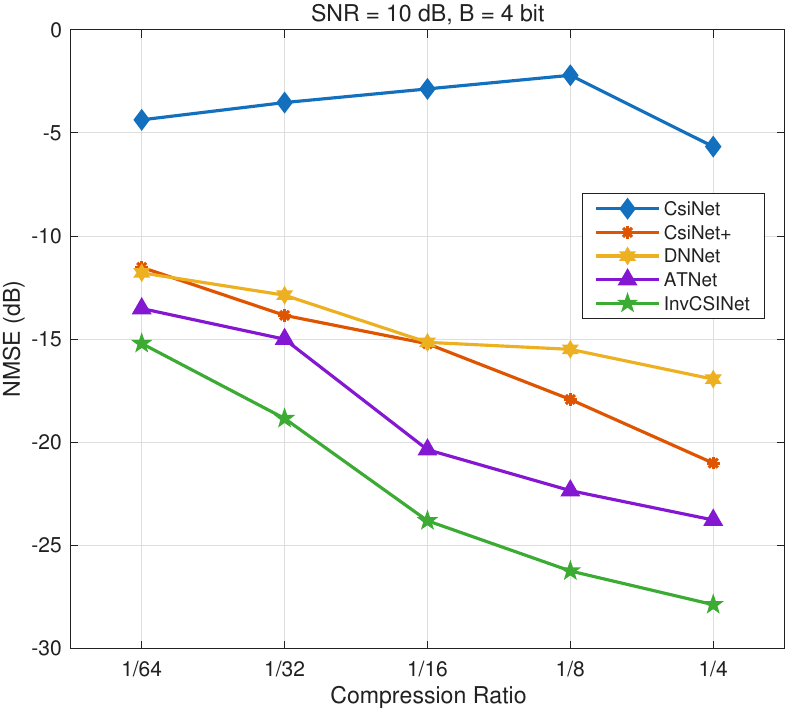}
\par\end{centering}
}

\caption{NMSE performance versus the compression ratio over the AWGN channel
at SNR = 0 dB (subfigures(a)--(d)) and SNR = 10 dB (subfigures (e)--(h)).\label{fig:NMSE-performance-versus}}
\end{figure*}

\subsection{Performance of INN-based CSI Feedback with Ideal Transmission}

In this subsection, we use CsiNet \cite{wen2018deep} as baseline
to evaluate the effectiveness of INN for CSI feedback and empirically
validate \textbf{Theorem} \ref{thm:pdf_match}. Two training scenarios
are considered: 1) using only the forward loss $\mathcal{L}_{\mathbf{r}}$
(i.e., the MMD term), and 2) using the combined loss $\mathcal{L}_{\mathbf{r}}+\kappa\cdot\mathcal{L}_{\mathbf{H}}$.

Fig. \ref{fig:NMSE-versus-the} and Fig. \ref{fig:MMD-versus-the}
respectively presents the NMSE and the MMD value computed between
the reconstructed $\widehat{\mathbf{H}}$ and the original CSI $\mathbf{H}$
versus training epochs. First, it shows that INN trained with either
the forward loss or the combined loss outperforms CsiNet. This performance
gain is attributed to the bijective nature of the INN architecture,
which facilitates information preservation during the compression
and reconstruction process, thereby enhancing recovery accuracy. In
contrast, the DAE-based CsiNet suffers from information loss, degrading
performance. Second, it is observed that training with only the distribution
matching loss $\mathcal{L}_{\mathbf{r}}$ is sufficient to achieve
strong distributional and sample-wise recovery consistency, verifying
the results in \textbf{Theorem} \ref{thm:pdf_match} and demonstrating
INN\textquoteright s potential for unsupervised learning. Third, as
shown in Fig. \ref{fig:MMD-versus-the}, it is hard for the MMD metric
to converge exactly to zero during training, indicating a residual
discrepancy between the prescribed and true distributions. In this
case, it is important to incorporate an additional MSE loss $\mathcal{L}_{\mathbf{H}}$
during training, which significantly boosts reconstruction accuracy
by penalizing mismatches in the signal domain.

\subsection{Performance of InvCSINet with Practical Impairments}

To verify the performance of proposed InvCSINet under practical transmission
conditions, Fig. \ref{fig:NMSE-performance-versus} presents the final
CSI reconstruction NMSE versus the compression dimension under different
SNR values and quantization bits. Compared with four representative
baselines, the proposed InvCSINet demonstrates consistently superior
performance under various simulation setups by endogenously incorporating
both quantization and bit-level channel distortion into the end-to-end
training process. Specifically, compared to CsiNet, which is based
on a DAE architecture and ignores quantization and channel errors,
InvCSINet achieves significant performance gains by preserving information
through its bijective structure and distortion-aware training. Compared
to CsiNet+, which employs a fixed $\mu$-law quantizer and an offline-trained
dequantizer but neglects the impact of channel distortion during training,
InvCSINet exhibits significant performance gains especially in low
SNR regime. Compared to DNNet, which incorporates a denoising module
to handle channel errors, InvCSINet still outperforms it across a
range of scenarios. While DNNet adopts the noise-aware design, its
lack of quantization-awareness leads to degraded overall performance,
and in some cases, results in unstable recovery. Compared to ATNet,
which considers both quantization and channel noise, InvCSINet achieves
better accuracy due to its bijective structure, fully differentiable
and unified training strategy. The two-stage optimization in ATNet,
caused by the non-differentiability of quantization and bit-level
errors, forces the gradients to be set to 1 and results in suboptimal
performance, especially under severe distortions.

\subsection{Ablation Study}

\begin{table*}[!t]
\caption{NMSE performance of InvCSINet w/o different modules.\label{tab:ablation_study}}

\centering{}%
\begin{tabular}{|c|c|c|c|c|}
\hline 
 & \multicolumn{4}{c|}{Neworks}\tabularnewline
\hline 
\multirow{2}{*}{Settings} & \multicolumn{3}{c|}{InvCsiNet w/o module} & \multirow{2}{*}{InvCsiNet}\tabularnewline
\cline{2-4}
 & DAQ & D-BCD & IC & \tabularnewline
\hline 
$\text{SNR}=0$ dB, $1/32$, $B=1$ bit & $-4.800$ dB & $-1.310$ dB & $-4.952$ dB & $\boldsymbol{-4.994}$ dB\tabularnewline
$\text{SNR}=0$ dB, $1/16$, $B=1$ bit & $-6.747$ dB & $-6.523$ dB & $-6.942$ dB & $\boldsymbol{-7.066}$ dB\tabularnewline
$\text{SNR}=0$ dB, $1/8$, $B=3$ bit & $-11.147$ dB & $-10.131$ dB & $-11.477$ dB & $\boldsymbol{-11.515}$ dB\tabularnewline
$\text{SNR}=10$ dB, $1/64$, $B=4$ bit & $-14.816$ dB & $-3.346$ dB & $-14.760$ dB & $\boldsymbol{-15.107}$ dB\tabularnewline
$\text{SNR}=10$ dB, $1/32$, $B=4$ bit & $-17.804$ dB & $-8.340$ dB & $-18.569$ dB & $\boldsymbol{-18.888}$ dB\tabularnewline
$\text{SNR}=10$ dB, $1/16$, $B=1$ bit & $-11.733$ dB & $-12.210$ dB & $-13.234$ dB & $\boldsymbol{-13.238}$ dB\tabularnewline
$\text{SNR}=10$ dB, $1/4$, $B=3$ bit & $-23.426$ dB & $-21.691$ dB & $-23.701$ dB & $\boldsymbol{-23.774}$ dB\tabularnewline
\hline 
\end{tabular} 
\end{table*}

\begin{table*}
\centering{}\caption{Comparison of parameters for different networks.\label{comparsion_params}}
\begin{tabular}{|c|c|c|c|c|c|c|c|c|}
\hline 
\multirow{1}{*}{} & \multicolumn{8}{c|}{Networks}\tabularnewline
\hline 
\multirow{2}{*}{Compress Ratio} & \multirow{2}{*}{CsiNet\cite{wen2018deep}} & \multirow{2}{*}{CsiNet+\cite{chen2019novel}} & \multirow{2}{*}{DNNet\cite{ye2020deep}} & \multirow{2}{*}{ATNet\cite{chang2021deep}} & \multicolumn{4}{c|}{InvCSINet}\tabularnewline
\cline{6-9}
 &  &  &  &  & $B=1$ bit & $B=2$ bit & $B=3$ bit & $B=4$ bit\tabularnewline
\hline 
1/64 & \textbf{136,560} & 145,870 & 1,252,816 & 160,592 & 292,452 & 292,580 & 292,836 & 293,348\tabularnewline
1/32 & \textbf{267,664} & 286,286 & 1,449,552 & 298,032 & 309,974 & 310,230 & 310,742 & 311,766\tabularnewline
1/16 & 529,872 & 585,550 & 1,843,024 & 585,200 & \textbf{351,978} & 352,490 & 353,514 & 355,562\tabularnewline
1/8 & 1,054,288 & 1,257,806 & 2,629,968 & 1,208,688 & \textbf{463,826} & 464,850 & 466,898 & 470,994\tabularnewline
1/4 & 2,103,120 & 2,897,230 & 4,203,856 & 2,652,272 & \textbf{798,882} & 800,930 & 805,026 & 813,218\tabularnewline
\hline 
\end{tabular}
\end{table*}

In the proposed InvCSINet, three key modules are designed to address
challenges encountered in practical CSI feedback transmission. To
evaluate the contribution of each module to the final CSI reconstruction
accuracy, we conduct a series of ablation studies. Table \ref{tab:ablation_study}
summarizes the NMSE performance changes for CSI reconstruction under
each ablation setting across different SNRs, compression dimensions,
and quantization bits.

\subsubsection{W/O DAQ Module}

The DAQ module is replaced with a uniform quantizer, while keeping
all other parameters and hyperparameters unchanged. This study evaluates
whether the learnable DAQ module improves CSI reconstruction accuracy
by adaptively designing the quantization strategy. It shows that removing
the DAQ module leads to a performance degradation of approximately
0.1 to 1.5 dB. The degradation becomes significant in high SNR regimes,
where channel distortion is minor and quantization distortion becomes
dominant. This suggests that when quantization becomes the primary
performance bottleneck, incorporating a DAQ module is essential to
mitigate quantization distortions and improve reconstruction accuracy.

\subsubsection{W/O D-BCD Module}

The D-BCD module is replaced by directly adding AWGN noise to the
discrete features $\mathbf{v}$ output from the DAQ module. All other
parameters and hyperparameters remain unchanged. The results demonstrate
the critical role of the D-BCD module. By modeling the bit-channel
distortion through a differentiable function and integrating it endogenously
into the network training, the CSI reconstruction accuracy can be
significantly improved. The absence of this module leads to a performance
degradation of approximately 0.5 to 11.8 dB, particularly notable
when the quantization bit is high (e.g., $B=4$ bits). This indicates
that when quantization distortion is relatively minor and channel
distortion dominates, it is crucial to incorporate the D-BCD module
during training to ensure robust performance.

\subsubsection{W/O IC Module}

The IC module is removed from InvCSINet, with all other parameters
and hyperparameters unchanged. The results show that removing the
IC module leads to a performance degradation of approximately 0.004
to 0.4 dB. By enforcing consistency between the output feature and
the original compressed representation, the IC module helps preserve
information throughout the entire feedback process. Although the performance
gain brought by the IC module is relatively modest, this is primarily
because most of the information loss induced by quantization and channel
distortion has already been captured and mitigated by the DAQ module
and the D-BCD model. Nonetheless, the IC module offers additional
flexibility to compensate for residual information loss across the
compression, quantization, and transmission stages.

\subsection{Model Complexity Comparison}

We compare the model sizes of considered schemes in terms of parameter
count, as summarized in Table \ref{comparsion_params}. The model
size of InvCSINet is relatively small and does not increase significantly
with the compression dimensions. In contrast, the four baseline methods
use completely separate parameter sets for the encoder and decoder,
leading to a rapid increase in parameter count with growing compression
dimensions. Overall, the proposed InvCSINet achieves superior performance
with fewer network parameters, demonstrating its parameter-efficiency
in practical deployments.

\section{Conclusion \label{sec_5} }

This paper proposes InvCSINet, a novel INN-based framework for information-preserving
CSI feedback in FDD massive MIMO systems. To address practical challenges
such as quantization and channel noise, InvCSINet integrates a DAQ
module, D-BCD module, and IC module, supporting end-to-end training
and robust CSI recovery. Theoretical analysis shows that, under perfect
training with zero convergence loss, InvCSINet preserves the CSI distribution
despite distortions caused by quantization and channel noise. Extensive
simulations on DeepMIMO verify its superior performance over baseline
methods, validating its practical effectiveness.

\appendices{}

\section{Proof of Theorem \ref{thm:pdf_match}}

\label{sec:Proof-of-thm:pdf_match}
\begin{IEEEproof}
If the MMD loss converges to zero, the network output follows the
prescribed distribution \cite{gretton2012kernel}:
\begin{equation}
q(\mathbf{z},\mathbf{r})=p(\mathbf{z})p_{R}(\mathbf{r}).
\end{equation}
Given an input channel $\mathbf{H}^{\star}$, sampled from a prior
distribution $p_{H}(\mathbf{H})$, we transform it using the forward
pass of perfected converged INN to obtain the distribution of output
$q^{\star}(\mathbf{z},\mathbf{r})$. Because $\mathcal{L}_{\mathbf{r}}=0$,
we have $q^{\star}(\mathbf{z},\mathbf{r})=p(\mathbf{z})p_{R}(\mathbf{r})$.
The marginal distribution of $\mathbf{z}$ must be $q^{\star}(\mathbf{z})=\int q^{\star}(\mathbf{z},\mathbf{r})\mathrm{d}\mathbf{r}=\delta(\mathbf{z}-f_{\mathbf{z}}(\mathbf{H}^{\star})).$
Also, because the independence between $\mathbf{z}$ and $\mathbf{r}$
in the output, the marginal distribution of $\mathbf{r}$ must be
$q^{\star}(\mathbf{r})=\int q^{\star}(\mathbf{z},\mathbf{r})\mathrm{d}\mathbf{z}=p_{R}(\mathbf{r})$.
So the joint distribution of outputs is
\begin{equation}
q^{\star}(\mathbf{z},\mathbf{r})=\delta\left(\mathbf{z}-f_{\mathbf{z}}(\mathbf{H}^{\star})\right)p_{R}(\mathbf{r}).
\end{equation}
Therefore, when we input $\mathbf{z}=f_{\mathbf{z}}(\mathbf{H}^{\star})$
back to the INN while sampling $\mathbf{r}$ according to $p_{R}$,
this is same as sampling $[\mathbf{z},\mathbf{r}]$ from the $q^{\star}(\mathbf{z},\mathbf{r})$
above. Based on Lemma \ref{lem:Ditribution-Preservation-of}, the
INN will output samples from $p_{H}$. That means the divergence between
the output distribution of the INN $q(\mathbf{H})$ and the prior
data distribution $p_{H}(\mathbf{H})$ will be 0.
\end{IEEEproof}

\section{Proof of Theorem \ref{thm:pdf_match-1}}

\label{sec:Proof-of-thm:pdf_match-1}
\begin{IEEEproof}
If the MMD loss converges to zero, the network output follows the
prescribed distribution \cite{gretton2012kernel}:
\begin{equation}
q(\mathbf{z},\mathbf{r})=p(\hat{\mathbf{z}})\upsilon_{R}(\mathbf{r}).\label{eq:joint_q}
\end{equation}
Given an input channel $\mathbf{H}^{\star}$, sampled from a prior
distribution $p_{H}(\mathbf{H})$, we transform it using the forward
pass of perfected converged InvCSINet to obtain the distribution of
output $q^{\star}(\mathbf{z},\mathbf{r})$. Because $\mathcal{L}_{\mathbf{r}}=0$,
we have 
\begin{align}
q^{\star}(\mathbf{z},\mathbf{r}) & =p(\hat{\mathbf{z}})\upsilon_{R}(\mathbf{r}).
\end{align}
The marginal distribution of $\mathbf{z}$ must be $q^{\star}(\mathbf{z})=\int q^{\star}(\mathbf{z},\mathbf{r})\mathrm{d}\mathbf{r}=\delta(\mathbf{z}-f_{\mathbf{z}}(\mathbf{H}^{\star})).$
Also, because the independence between $\mathbf{z}$ and $\mathbf{r}$
in the output, the marginal distribution of $\mathbf{r}$ must be
$q^{\star}(\mathbf{r})=\upsilon_{R}(\mathbf{r})$. So the joint distribution
of outputs is 
\begin{equation}
q^{\star}(\mathbf{z},\mathbf{r})=\delta\left(\mathbf{z}-f_{\mathbf{z}}(\mathbf{H}^{\star})\right)\upsilon_{R}(\mathbf{r}).\label{eq: q}
\end{equation}
Connecting with \eqref{eq:joint_q}, we have
\begin{equation}
\delta\left(\hat{\mathbf{z}}-f_{\mathbf{z}}(\mathbf{H}^{\star})\right)=p(\hat{\mathbf{z}}).\label{eq:z_hat=00003Dz}
\end{equation}
Since $\hat{\mathbf{z}}$ is obtained by passing $\mathbf{z}^{\star}=f_{\mathbf{z}}(\mathbf{H}^{\star})$
through the perfected converged DAQ module, D-BCD module and IC module,
we have $\hat{\mathbf{z}}^{\star}=\varPsi\left(\Omega\left(\Lambda^{\mathrm{DAQ}}\left(f_{\mathbf{z}}(\mathbf{H}^{\star})\right),\bar{\mathbf{V}}\right)\right)$.
As a result, the marginal distribution of $\hat{\mathbf{z}}$ at the
input of INN inverse path satisfies
\begin{align}
p(\hat{\mathbf{z}}) & =\delta(\hat{\mathbf{z}}-\hat{\mathbf{z}}^{\star})\nonumber \\
 & =\delta\left(\hat{\mathbf{z}}-\varPsi\left(\Omega\left(\Lambda^{\mathrm{DAQ}}\left(f_{\mathbf{z}}(\mathbf{H}^{\star})\right),\bar{\mathbf{V}}\right)\right)\right).\label{eq:z_hat=00003Dz_star}
\end{align}
From \eqref{eq:z_hat=00003Dz} and \eqref{eq:z_hat=00003Dz_star},
the following equality between $\mathbf{z}^{\star}$ and $\hat{\mathbf{z}}^{\star}$
can be achieved:
\begin{equation}
\hat{\mathbf{z}}^{\star}=\varPsi\left(\Omega\left(\Lambda^{\mathrm{DAQ}}\left(f_{\mathbf{z}}(\mathbf{H}^{\star})\right),\bar{\mathbf{V}}\right)\right)=f_{\mathbf{z}}(\mathbf{H}^{\star}).\label{eq: equality}
\end{equation}
Therefore, when we input $\hat{\mathbf{z}}^{\star}=f_{\mathbf{z}}(\mathbf{H}^{\star})$
back to the INN while sampling $\mathbf{r}$ according to $\upsilon_{R}$,
this is same as sampling $[\mathbf{z},\mathbf{r}]$ from the $q^{\star}(\mathbf{z},\mathbf{r})$
in \eqref{eq: q}. Based on Lemma \ref{lem:Ditribution-Preservation-of},
the INN will output samples from $p_{H}$. That means the divergence
between the outputs of the INN $q(\mathbf{H})$ and the prior data
distribution $p_{H}(\mathbf{H})$ will be 0.
\end{IEEEproof}
\bibliographystyle{IEEEtran}
\bibliography{IEEEabrv,InvCSINet}

\end{document}